\documentclass[aps,pra,amssymb,superscriptaddress,twocolumn,notitlepage,nofootinbib,longbibliography,floatfix]{revtex4-2}

\usepackage[final]{graphicx}
\usepackage{times,bbm,amsmath,amssymb,amsthm}
\usepackage{epsfig,color}
\usepackage{xcolor}
\usepackage[colorlinks=true,citecolor=blue,linkcolor=blue]{hyperref}
\hypersetup{
    allcolors  = {blue},
}

\usepackage{cleveref}

\usepackage[caption = false]{subfig}
\usepackage{verbatim}
\usepackage[greek,english]{babel}
\usepackage{thumbpdf,enumerate}
\usepackage{booktabs}
\usepackage{sidecap}
\usepackage[scaled=.8]{couriers}    
\usepackage{pstricks}
\usepackage{multirow}
\usepackage{placeins}
\usepackage{relsize}  
\usepackage{pst-grad,bm}
\usepackage{epigraph}
\usepackage{longtable}
\usepackage{booktabs}
\usepackage{dsfont}
\usepackage{mathtools}

\usepackage{qcircuit}

\usepackage{soul}
\usepackage{ulem} 
\usepackage{acronym}
\usepackage{physics}
\usepackage{tikz}

\begin{document}

\title{Complexity and multi-functional variants of the Quantum-to-Quantum Bernoulli Factories}

\author{Francesco Hoch} 
\affiliation{Dipartimento di Fisica, Sapienza Universit\`{a} di Roma,
Piazzale Aldo Moro 5, I-00185 Roma, Italy}

\author{Taira Giordani} 
\affiliation{Dipartimento di Fisica, Sapienza Universit\`{a} di Roma,
Piazzale Aldo Moro 5, I-00185 Roma, Italy}

\author{Gonzalo Carvacho}
\affiliation{Dipartimento di Fisica, Sapienza Universit\`{a} di Roma,
Piazzale Aldo Moro 5, I-00185 Roma, Italy}

\author{Nicol\`o Spagnolo}
\affiliation{Dipartimento di Fisica, Sapienza Universit\`{a} di Roma,
Piazzale Aldo Moro 5, I-00185 Roma, Italy}

\author{Fabio Sciarrino}
\email[Corresponding author: ]{fabio.sciarrino@uniroma1.it}
\affiliation{Dipartimento di Fisica, Sapienza Universit\`{a} di Roma,
Piazzale Aldo Moro 5, I-00185 Roma, Italy}

\begin{abstract}
A Bernoulli factory is a model for randomness manipulation that transforms an initial Bernoulli random variable into another Bernoulli variable by applying a predetermined function relating the output bias to the input one. In literature, quantum-to-quantum Bernoulli factory schemes have been proposed, which encode both the input and output variables using qubit amplitudes. This fundamental concept can serve as a subroutine for quantum algorithms that involve Bayesian inference and Monte Carlo methods, or that require data encryption, like in blind quantum computation. In this work, we present a characterisation of the complexity of the quantum-to-quantum Bernoulli factory by providing a lower bound on the required number of qubits needed to implement the protocol, an upper bound on the success probability and the quantum circuit that saturates the bounds. We also formalise and analyse two different variants of the original problem that address the possibility of increasing the number of input biases or the number of functions implemented by the quantum-to-quantum Bernoulli factory. The obtained results can be used as a framework for randomness manipulation via such an approach.

\end{abstract}

\maketitle
\section{Introduction}

A Bernoulli factory is a paradigm of randomness processing first proposed by Keane and O’Brien \cite{Keane1994} to manipulate Bernoulli distributions only from their sampling.
More formally, a Bernoulli factory is an algorithm that, given independent samples from a Bernoulli distribution (a coin) with unknown bias $p$, exactly simulates a Bernoulli variable with bias $q = f(p)$ where $f: S \subseteq [0,1] \rightarrow [0,1]$ is a predetermined function called classical-to-classical Bernoulli factory (CCBF).
Starting from the original problem \cite{PaesLeme2023, Niazadeh2024, Niazadeh2023}, various applications and variants \cite{flegal2012exact, vats2022efficient, gonccalves2023exact, herbei2014estimating, Koskela2023, Morina2022} have been proposed over time. In particular, algorithms that implement the protocol for specific classes of functions \cite{Goyal2012, Mossel2005, atuszyski2011, Nacu2005, Mendo2019} and an analysis of the computational complexity measured in terms of the average number of samples \cite{Flajolet2011, Mendo2019, Holtz2010} have been addressed.

Due to the generality of the Bernoulli factory problem and the intrinsic stochastic nature of quantum mechanics, quantum extensions of the problem have been introduced.
The first quantum extension was addressed by Dale et al. \cite{Dale2015, Dale2016} in which the input classical coin is replaced with its corresponding quantum version. This problem shows interesting properties since it has a larger set of simulable functions and there is evidence of an advantage in terms of the average number of samples needed for the algorithm compared to the classical counterpart \cite{Patel2019}. This version is called quantum-to-classical Bernoulli factory (QCBF) and has been experimentally implemented on superconducting \cite{yuan2016} and photonic \cite{Patel2019} platforms.
Subsequently, Jiang et al. \cite{Jiang2018} formalise a new version called Quantum-to-quantum Bernoulli factory (QQBF) in which the entire Bernoulli process is quantum, i.e. both the input and output states.
Moreover, they prove that a complex function $f(z)$ is simulable by a QQBF if and only if it is a complex rational function. This proof was obtained by exploiting the field properties of the set of rational functions. In this regard, experimental implementations through photonic platforms have been devised \cite{Zhan2020, liu2021general, Hoch2024, Rodari2024}, opening the path to the application of the Bernoulli factory as a subroutine for quantum algorithms such as the blind quantum computation \cite{Kashefi2017, Fitzsimons2017, Polacchi2023}.

Despite theoretical and experimental advances, how the complexity of the QQBF problem, in terms of qubit count and maximum success probability, depends on the implemented function remains an open question.
In this manuscript, we present a refined demonstration to characterise the set of simulable functions, offering significantly different perspectives over the existing literature \cite{Jiang2018}. While confirming the prior results, our approach directly yields new insights into the protocol resources requirements, specifically, the number of qubits and the maximal success probability. Furthermore, it enables the explicit construction of a quantum circuit with optimal performance. 
In addition to the complexity analysis of the QQBF, two different extensions of the original problem are formalised and analysed. The first one is called \textit{Multivariate quantum-to-quantum Bernoulli factory} and extends the QQBF problem to the possibility of having more quantum Bernoulli distributions with different biases at the input.
The second extension introduces the \textit{Multifunctional Quantum-to-Quantum Bernoulli Factory}, which formalises the implementation of multiple functions within a single QQBF. While prior experimental demonstrations have featured gates that can be interpreted in this context \cite{Hoch2024, Rodari2024, Jiang2018}, a general theoretical framework for this class of problems has not been devised until now.

The work is organised as follows. 
In Sec.~II, we provide some definitions and concepts at the basis of this study. 
Sec.~III focuses on the demonstrations of the complexity of a quantum-to-quantum Bernoulli factory as well as on the analysis of the results. 
In Sec.~IV and Sec.~V, we formalise and analyse the two new extensions of the original problem, respectively the multivariate quantum-to-quantum Bernoulli factory and the multifunctional quantum-to-quantum Bernoulli factory. 
In Sec.~VI we present an application of the theory 
for the implementation of two elementary QQBF operations, such as the sum and the product between quantum states.
Finally, in Sec.~VII we summarise the results and discuss the possible applications and directions opened by our analysis.

\section{Background and definitions}

Quantum-to-quantum Bernoulli factories (QQBFs) operate on the amplitudes of qubit states. 
In Refs. \cite{Dale2015, Dale2016} the concept of a quantum coin (or quoin) defined as a qubit in a pure state $\ket{C_p} = \sqrt{p} \ket{0_c}+ \sqrt{1-p} \ket{1_c}$ was introduced.
This definition is motivated by the fact that, if the state is measured on the computational basis, it returns a classical Bernoulli variable with bias $p$.
In this work, we use the more general parametrisation of Refs.~\cite{Hoch2024,Rodari2024} in which each qubit state is associated with a complex number $z \in \mathds{C}$ through the expression
\begin{equation}
    \ket{z} \coloneqq \frac{z \ket{0_c} + \ket{1_c}}{\sqrt{1+\abs{z}^2}},
\end{equation}
where $\ket{0_c}$ and $\ket{1_c}$ are the two vectors of the computational basis.
This parameterisation is related to the bijective map associated with the stereographic projection that links each point in the complex plane to a point on the surface of the Bloch sphere.
In this way, we can identify each quoin with the state $\ket{C_p} = \ket{\sqrt{\frac{p}{1-p}}}$ and extend the concept to all the Bloch sphere.

In this framework, a quantum-to-quantum Bernoulli factory associated with a complex function ${f(z): \mathds{C} \rightarrow \mathds{C}}$ is the protocol that takes as input $n$ qubits in the state $\ket{z}$ and returns one qubit in the state $\ket{f(z)}$ with a certain non-null success probability.
In this contest, we say that a complex function ${f(z): \mathds{C}  \rightarrow \mathds{C}}$ is \textit{simulable} if there exists a QQBF that implements it.
It is known from Ref. \cite{Jiang2018} that the set of simulable functions by a QQBF is the set of \emph{complex rational functions}.

A complex rational function $r(z) \in \text{Rat}(z)$ is defined as the ratio of two polynomials $P(z)$ and $Q(z)$ with complex coefficients
\begin{equation}
    r(z) = \frac{P(z)}{Q(z)},
\end{equation}
where $Q(z)$ is a polynomial different from zero and $P(z)$ and $Q(z)$ have no common factor. Furthermore, we define the degree of a complex rational function as the maximum between the degrees of the two polynomials 
$\deg(r(z)) = \max\{\deg(P), \deg(Q)\}$.

 We can extend the concept by defining the \textit{multivariate complex rational functions} as 
\begin{equation}
    s(z_1, \dots, z_k) = \frac{P(z_1, \dots, z_k)}{Q(z_1, \dots, z_k)},
\end{equation}
where $s(z_1, \dots, z_k)\in \text{Rat}(z_1, \dots, z_k)$ is a function of $k$ variables   and  $P(z_1, \dots, z_k)$ and $Q(z_1, \dots, z_k)$ are two polynomials with complex coefficients. Similarly to the single variable case, the degree of the function related to the variable $z_j$ is defined as the maximum between the degrees of the two polynomials in that particular variable 
$\deg_j(s(z_1, \dots, z_k)) = \max\{\deg_j(P), \deg_j(Q)\}$.

\section{Quantum-to-quantum Bernoulli factory complexity}
\label{sec:QQBF}

In the following, we provide a formal derivation of the class of simulable functions by a QQBF and the quantum resources needed to implement a given function. Our demonstrations do not assume any particular internal structure of the QQBF. This contrasts with previous demonstrations in Refs. \cite{Jiang2018} where the authors individuated the class of simulable functions assuming the fundamental operations of the complex field, i.e. sum, inversion, and product. We show that such a structure for the QQBF can be retrieved as a special case of the demonstration presented in this work. 
Furthermore, such a characterisation of the set of simulable functions allows the derivation of bounds for the problem complexity in terms of the minimum number of qubits required for the implementation and the maximum success probability.
Finally, it is also shown how to construct the quantum circuit for a given function that allows the saturation of the bounds.

\subsection{Generic circuit representation of the QQBF: superset of simulable functions and lower bound for the number of qubits}

\begin{figure}[ht!]
        \centering
        \[\Qcircuit @C=1em @R=1em {
        && \lstick{\ket{z}} & \qw & \multigate{6}{U} & \qw & \qw &\rstick{\ket{\psi_O}} \qw\\
         && \lstick{\ket{z}} & \qw & \ghost{U} & \qw & \meter & \rstick{q_2 = 0} \cw\\
        && \cdots& & \nghost{U} & & \cdots \\
        && \lstick{\ket{z}} & \qw & \ghost{U} & \qw & \meter & \rstick{q_n = 0} \cw\\
        && \lstick{\ket{0_c}} & \qw & \ghost{U} & \qw & \meter & \rstick{q_{n+1} = 0} \cw\\
        && \cdots& & \nghost{U} & & \cdots \\
        && \lstick{\ket{0_c}} & \qw & \ghost{U} & \qw & \meter & \rstick{q_{n+m} = 0} \cw\\
        {\inputgroupv{1}{4}{1em}{3.2 em}{n}} \\
        {\inputgroupv{5}{7}{1em}{2.1 em}{m}} \\
        }\]
        
        \caption{\textbf{Schematic representation of the general circuit for the Quantum Bernoulli.} The circuit takes as input $n$ qubit in the state $\ket{z}$ and $m$ ancillary qubits in the state $\ket{0_c}$. The resulting state evolves through a unitary evolution $U$. All qubits are then measured in the computational basis, except the first one, which carries the output state of the Bernoulli factory. The output state $\ket{\psi_O}$ is accepted if all the measurements $q_2 \dots q_{n+m}$ return the outcome $0$.
        This circuit is the most general one since every circuit implementing a Quantum-to-quantum Bernoulli factory can be mapped into the one shown in the figure, as described in the main text.}
        \label{fig: BF_circuit}
\end{figure}
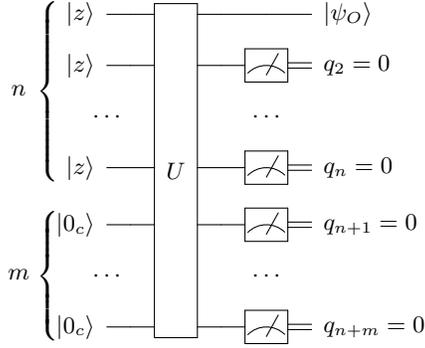

The first step to characterise the set of simulable functions is to show that every circuit implementing a Quantum-to-quantum Bernoulli factory can be mapped into a circuit in the form of Fig.~\ref{fig: BF_circuit}.
Since the hidden information in a Bernoulli factory circuit is provided only by the unknown state $\ket{z}$ we can divide the circuit into three principal components.
First, the input state, which is composed of $n$ copies of the state $\ket{z}$ and some ancillary qubits in known pure states. Then, a unitary evolution and a measurement are performed, such that at the end only a single qubit remains that encodes the output state. Using the deferred measurement principle \cite{Nielsen2009, Gurevich2021} we can postpone all the measurements at the end of the circuit by eventually adding more ancillary qubits. In this way, we have only one known global unitary evolution between the initial state and the measurement. Furthermore, without loss of generality, we can assume that all ancillary qubits are in the state $\ket{0_c}$ since any other state can be generated by applying a unitary transformation which can be embedded in the global one. With a similar argument, we can say that all measurements are performed in the computational basis and that the result is accepted when all of them provide the outcome $0$.

Using such a framework, we can identify the superset of the simulable functions, i.e. a set that contains the set of simulable functions and the lower bound on the required number of qubits to simulate a given function.

First, we write the input state of the circuit in Fig~\ref{fig: BF_circuit} as 
\begin{equation}
\begin{split}
    \ket{\psi_I} &= \ket{0_c}^{\otimes m} \otimes \biggl(\frac{z\ket{0_c}+\ket{1_c}}{\sqrt{1+\abs{z}^2}}\biggr)^{\otimes n} \\ &= \biggl(\frac{1}{1+\abs{z}^2}\biggr)^{\frac{n}{2}}\sum_{j=0}^n z^j  \sqrt{\binom{n}{j}} \ket{0_c}^{\otimes m} \otimes  \ket{s_j^n},
\end{split}
\end{equation}
where $\ket{s_j^n}$ is the totally symmetric state of $n$ qubits with $j$ zeros. It is important to note that the coefficients composing the state are monomial in $z$ and span all the powers from $0$ to $n$. 
Since the unitary evolution is linear, then all the coefficients of the final state before the measurements are a linear combination of the input state coefficients.
\begin{equation}
    \ket{\psi} = \frac{1}{\left(1+\abs{z}^2\right)^{\frac{n}{2}}} \displaystyle \sum_{k = 0}^{2^{n+m}} P_k(z) \ket{k},
\end{equation}
where
\begin{equation}
    P_k(z) = \sum_{j=0}^n \sqrt{\binom{n}{j}} u_{kj} z^j,
\end{equation}
and $\ket{k}$ is the state represented by the bitstring of the number $k$ and $u_{kj} = \bra{k}U\left(\ket{s_j^n}\otimes\ket{0}^{\otimes m}\right)$.
This consideration demonstrates that all the coefficients of the state written in the computational basis are polynomial in $z$ with the degree at most $n$. At this point, we apply the measurements. Since all are performed on the computational basis and the results must be all $0$, this means that we are selecting the states $\ket{k=0}$ and $\ket{k=1}$ and that the output state of such a QQBF is in the form:
\begin{equation}
    \ket{\psi_O} = \frac{P(z) \ket{0_c}+ Q(z) \ket{1_c}}{\sqrt{\abs{P(z)}^2+\abs{Q(z)}^2}} = \ket{f(z) = \frac{P(z)}{Q(z)}},
\end{equation}
where $P(z) = P_0(z)$ and $Q(z) = P_1(z)$. 
We have then retrieved the same result of Ref. \cite{Jiang2018} that the space of simulable functions $S$ is a subset of the set of rational functions $S \subseteq \text{Rat}(z)$.
Most importantly, we note that in such a formulation of the problem, there is a direct relationship between the degree of the implemented function $f(z)$ and the number of resources. The functions implementable by a circuit with $n$ copies of the state $\ket{z}$ as input have a degree of $n$ at most.  This result allows us to have a lower bound on the minimum number of qubits in the state $\ket{z}$ required to implement a given function so as the total number of qubits required. Indeed, the optimal QQBF in terms of resources requires at least as many qubits as the degree of the function:
\begin{equation}
    n\geq \deg(f(z)).
    \label{eq:bound_qubits}
\end{equation}

\subsection{Tightness of the bound and choice of the optimal unitary transformation}

Now that we found an upper bound for the set of simulable functions our goal is to prove that it is indeed an equality by showing that $\text{Rat}(z) \subseteq S$.
We also want to show that the lower bound of Eq.~\eqref{eq:bound_qubits} on the required number of bits is tight and no ancillary qubits are required ($m = 0$) except for the particular case of $n= 1$.

To accomplish this, for each tuple $(P(z), Q(z)) \in \mathcal{P}(\mathds{C}) \cross \mathcal{P}(\mathds{C})$ we construct a circuit implementing a Bernoulli factory with an associated function $f(z) = P(z)/Q(z)$ that use a number of qubits $n$ equal to the lower bound of eq. \eqref{eq:bound_qubits} and do not require ancillary qubits.
First, we describe the state at the output of the circuit in Fig.~\ref{fig: BF_circuit}:
\begin{equation}
    \ket{\psi_O} = \frac{\ket{0_c} \bra{v_0} \ket{\psi_I}+\ket{1_c} \bra{v_1} \ket{\psi_I}}{\sqrt{\abs{\bra{v_0} \ket{\psi_I}}^2+\abs{\bra{v_1} \ket{\psi_I}}^2}},
    \label{eq: Out_base_2}
\end{equation}
where we describe the two rows of the unitary matrix with two covectors $\bra{v_0} = \bra{0_c}^{\otimes n-1} \otimes \bra{0_c} U$ and ${\bra{v_1} = \bra{0_c}^{\otimes n-1} \otimes \bra{1_c} U}$.
Since we aim at implementing a Bernoulli factory with the associated function $g(z) = P(z)/Q(z)$, we impose the following equalities:
\begin{eqnarray}
   \bra{v_0} \ket{\psi_I} &=& K P(z), \\  
   \bra{v_1} \ket{\psi_I} &=& K Q(z),
\end{eqnarray}
for some real non-null proportional factor $K$. Defining $P(z) = \sum p_j z^j$, $Q(z) = \sum q_j z^j$, and using the property that the monomials with different powers are linearly independent we can rewrite the previous equations in the form:
\begin{eqnarray}
    \bra{v_0} \ket{s_j^n} =K \frac{p_j}{\sqrt{\binom{n}{j}}}, \\ 
    \bra{v_1} \ket{s_j^n} =K \frac{q_j}{\sqrt{\binom{n}{j}}}.
\end{eqnarray}

Using the property that the vectors $\ket{s_j^n}$ are orthogonal to each other we can write the two vectors $\ket{v_0}$ and $\ket{v_1}$ in the general form:

\begin{equation}
\begin{aligned}
    \ket{v_0} &= \sum_{j = 0}^{n} K \frac{p^*_j}{\sqrt{\binom{n}{j}}} \ket{s_j^n} + K x \ket{\theta_0}, \\ 
    \ket{v_1} &= \sum_{j = 0}^{n} K \frac{q^*_j}{\sqrt{\binom{n}{j}}} \ket{s_j^n} + K y \ket{\theta_0} + K w \ket{\theta_1},
\end{aligned}
\end{equation}
where $\ket{\theta_0}$ and $\ket{\theta_1}$ are two vectors orthogonal to each other and to all vectors $\ket{s_j^n}$, and $x$, $y$, $w$ are free parameters (for more details on the derivation see the Supplementary Materials I).
Without loss of generality, we can assume that $x$ and $w$ are real positive parameters since the eventual phases can be absorbed by a redefinition of the vectors $\ket{\theta_0}$ and $\ket{\theta_1}$.
Now we have to impose that the covectors associated with the vectors $\ket{v_0}$ and $\ket{v_1}$ are proper rows of a unitary transformation. Thus, we need to impose that the two vectors are orthonormal as follows:
\begin{eqnarray}
        \braket{v_0}{v_0} &=& 1 \Rightarrow \sum_{j = 0}^n \frac{\abs{p_j}^2}{\binom{n}{j}} + \abs{x} ^ 2 = 1/K^2, \\
        \braket{v_1}{v_1} &=& 1 \Rightarrow \sum_{j = 0}^n \frac{\abs{q_j}^2}{\binom{n}{j}} + \abs{y} ^ 2 + \abs{w} ^ 2 = 1/K^2,\\
        \braket{v_0}{v_1} &=& 0 \Rightarrow x^*y  =  -\sum_{j = 0}^n \frac{p_j q_j^*}{\binom{n}{j}}.
\end{eqnarray}
Considering that there are four variables but only three equations, we solve the system as a function of the free parameter $w$.
We first define the following variables: 
\begin{eqnarray}
        a &=& \sum_{j = 0}^n \frac{\abs{q_j}^2}{\binom{n}{j}}, \\ 
        b &=& \sum_{j = 0}^n \frac{\abs{p_j}^2}{\binom{n}{j}}, \\
        c &=& \sum_{j = 0}^n \frac{p_j q_j^*}{\binom{n}{j}}.
\end{eqnarray}
We can then solve the system of equations, leading to:
\begin{eqnarray}
        x &=&\sqrt{\frac{l+\abs{w}^2+a-b}{2}},
        \label{eq: x_y_variable_1}\\
         y &=&\frac{-c}{\abs{c}}\sqrt{\frac{l-\abs{w}^2-a+b}{2}}, \label{eq: x_y_variable_2}\\
         K &=&  \sqrt{\frac{2}{l+\abs{w}^2+a+b}},
        \label{eq: x_y_variable_3}
\end{eqnarray}
where $l=\sqrt{(\abs{w}^2+a-b)^2+4\abs{c}^2}$.

It is possible to note that the solution exists for every combination of the polynomials coefficients and for every value of $w$. Hence, to reduce the number of required basis vectors we fix $w = 0$ making the vector $\ket{\theta_1}$ unnecessary.

With those parameters, we can construct the orthonormal vectors $\ket{v_{0/1}}$ and by completing it to an orthonormal basis (see Supplementary Materials II) we can construct the corresponding unitary matrix $U$. The only remaining open issue is the possibility of finding a vector  $\ket{\theta_0}$ that is orthogonal to all the vectors $\ket{s_i^n}$, eventually with the need to use ancillary qubits to increase the Hilbert space dimension. A Hilbert space generated by $n$ qubits has dimension $2^n$ and the number of vectors $\ket{s_i^n}$ is $n+1$. Hence, if $2^n-n-1 > 0$ then there is at least a vector that can be used for $\ket{\theta_0}$ without the need to use ancillary qubits. The inequality is true for $n\geq 2$. This means that it is necessary to use an ancillary qubit only for certain functions of degree one or less. In the other cases the required number of qubits is equal to the lower bound of Eq. \eqref{eq:bound_qubits}.
We note also that $n=1$ there are functions where the ancillary qubit is not required. This happens when $x=0$ and $y = 0$ which translates to the condition $a = b$ and $c= 0$. 

This analysis proves that the set of simulable functions coincides with the set of rational functions $S = \text{Rat}(z)$. We also show the connection between the minimum number of qubits required to simulate a function and the degree of the function itself.

\subsection{Success probability}

Now we analyse the success probability of the quantum circuit to generate the correct output state. The success probability is
\begin{equation}
    \mathrm{Pr}(z) = \abs{\bra{v_0} \ket{\psi_I}}^2+\abs{\bra{v_1} \ket{\psi_I}}^2.
\end{equation}
By using the expression obtained previously, we can write
\begin{equation}
    \mathrm{Pr}(z) = \frac{K^2\biggl(\abs{\sum_{j=0}^n p_j z^j}^2+\abs{\sum_{j=0}^n q_j z^j}^2\biggr)}{\biggl(1+\abs{z}^2\biggr)^n}.
\end{equation}

\textbf{Optimisation over the parameter $w$.} First of all, we aim at finding the maximum success probability for a given number of exploited qubits (i.e. for fixed $n$). We maximise the success probability as a function of $\abs{w}^2$, since it is the only free parameter. Calculating the derivative, we can see that the success probability $\mathrm{Pr}(z)$ is always a decreasing function of $\abs{w}^2$, independently of other parameters. Hence, the maximum probability is reached for $\abs{w}^2 = 0$ showing that the previous construction of the vectors $\ket{v_{0/1}}$ is optimal for both the number of basis vectors and the success probability.
Hence, the maximum success probability $\mathrm{Pr}_n (z)$ for a given number of qubits $n$ is
\begin{equation}
    \mathrm{Pr}_n(z) = \frac{2\biggl(\abs{\sum_{j=0}^n p_j z^j}^2+\abs{\sum_{j=0}^n q_j z^j}^2\biggr)}{\biggl(1+\abs{z}^2\biggr)^n\left[\sqrt{\biggl(a-b \biggr)^2+4\abs{c}^2}+a+b\right]}.
\end{equation}

\textbf{Optimisation over $n$.} %
The next step is to find the maximum success probability by varying the number of qubits. A direct comparison between the success probabilities associated with two QQBF processes with a different number of qubits is not straightforward, since they depend on the states. Consequently, a well-defined optimization can be carried out by considering an input state ensemble and its probability distribution. To provide concrete insights, we therefore derive general bounds based on state ensemble analysis and complement them with a detailed examination of representative cases.

In general, if we use a number of qubits equal to the degree of the function $f(z)$, then the probability of success is non-zero for any value of $z$ (also for $z \rightarrow \infty$). This property follows from the fact that the polynomials $P(z)$ and $Q(z)$ that compose the function $f(z)$ are coprime and thus possess no zeros in common.
If we are interested in maximising the success probability for a particular state $\ket{z}$ or of its mean value over a state ensemble, then we can consider the possibility of increasing the number of qubits employed and therefore the degree of the polynomials. To accomplish this, given the function $f(z) = P(z)/Q(z)$, we multiply both polynomials $P(z)$ and $Q(z)$ by the same factor $z-r$, which modifies the success probability, increasing it for some value and adding a zero for $z = r$.
In this context, increasing only the number of employed qubits, leaving the polynomials unchanged, is equivalent to the case where $r \rightarrow \infty$.

As explicative examples, we analyze the case of the average success probability over the uniform distribution of all the qubits on the Bloch sphere. In such a scenario, the mean success probability will be
\begin{equation}
    \langle \mathrm{Pr}_n \rangle_U = \frac{2(a+b)}{(n+1)\left[\sqrt{\biggl(a-b \biggr)^2+4\abs{c}^2}+a+b\right]}.
\end{equation}
Instead, if we consider a uniform distribution only on the {covariant states}, i.e. states in the form $(e^{i\phi} \ket{0}+\ket{1})/\sqrt{2}$, then the mean success probability is: 
\begin{equation}
    \langle \mathrm{Pr}_n \rangle_C = \frac{ \sum_{j=0}^n \abs{p_j}^2 + \abs{q_j}^2}{2^n \left[\sqrt{\biggl(a-b \biggr)^2+4\abs{c}^2}+a+b\right]}.
\end{equation}

\begin{figure}[ht!]
    \centering
    \includegraphics[width = \linewidth ]{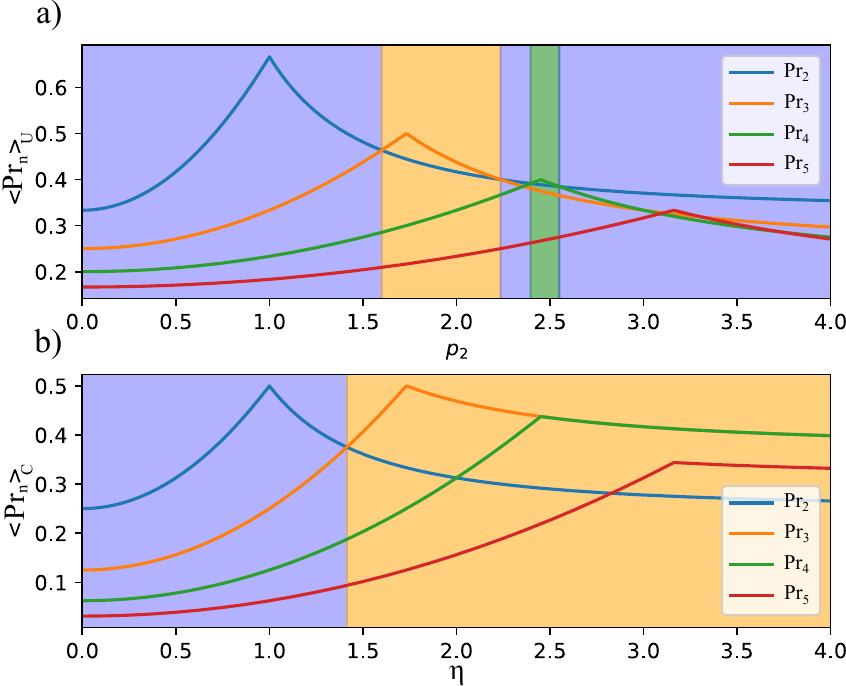}
    \caption{\textbf{Average success probabilities $\langle \mathrm{Pr}_n\rangle_{U/C}$ for the function $g(z) = \eta z^2$.} Representation of the success probability for the function $g(z) = \eta z^2$ for different numbers of qubits employed by the quantum circuit and for different sets of states. In a) we considered the average over the uniform distribution of all the qubits and in b) we consider the average only on the equatorial states. The background colour represents the region in which a particular probability is higher than the others.
    It can be deduced from the figures that the optimal number of qubits required depends strongly on both the function itself and the set of states considered. }
    \label{fig:Probability_function}
\end{figure}

Fig.~\ref{fig:Probability_function} shows the trends of those mean probabilities for the function $g(z) = \eta z^2$ as a function of the parameter $\eta$ and for different numbers of qubits $n$ employed by the quantum circuit of the QQBF. 
We note that, as the parameter $\eta$ changes, the number of optimal qubits changes in a non-trivial manner, and the regions are different depending on the set of states considered.

With these last considerations on success probabilities, we have fully characterised the Quantum-to-quantum Bernoulli factory problem. In summary, we have identified the set of simulable functions by showing that it coincides with the set of rational functions $S = \text{Rat}(z)$. We also show that the minimum number of qubits required for the simulation of a function $f(z) \in \text{Rat}(z)$ is $n = \deg(f(z))$ and no ancillary qubit is required ($m=0$), except in the case $\deg(f(z)) = 1$ where at most one is needed. In addition, a systematic procedure for constructing the quantum circuit related to a particular function was presented, showing how the circuit is the most efficient in terms of qubits used and success probability.
In this analysis, the possibility of tracing some qubits instead of measuring them was not discussed. In the Supplementary Materials III, we show that this possibility does not bring any advantage in terms of resources or success probability.  

\section{Multivariate Quantum-to-quantum Bernoulli factory}
\begin{figure}[ht!]
        \centering
        \[\Qcircuit @C=1em @R=1em {
        && \lstick{\ket{z_1}} & \qw & \multigate{9}{U} & \qw & \qw &\rstick{\ket{\psi_O}} \qw\\
        && \cdots& & \nghost{U} & \qw & \meter & \rstick{q_2 = 0} \cw\\
        && \lstick{\ket{z_1}} & \qw &\ghost{U} & \qw & \meter & \rstick{q_3 = 0} \cw\\
        && \cdots& & \nghost{U} & \qw & \meter & \rstick{q_4 = 0} \cw\\
        && \lstick{\ket{z_k}} & \qw & \ghost{U} \\
        && \cdots& & \nghost{U} && \cdots\\
        && \lstick{\ket{z_k}} & \qw &\ghost{U} \\
        && \lstick{\ket{0}} & \qw & \ghost{U} & \qw & \meter & \rstick{q_{n_t+m-2} = 0} \cw\\
        && \cdots& & \nghost{U} & \qw & \meter & \rstick{q_{n_t+m-1} = 0} \cw\\
        && \lstick{\ket{0}} & \qw & \ghost{U} & \qw & \meter & \rstick{q_{n_t+m} = 0} \cw\\
        {\inputgroupv{1}{3}{1em}{2.1 em}{n_1}} \\
        {\inputgroupv{5}{7}{1em}{2.1 em}{n_k}} \\
        {\inputgroupv{8}{10}{1em}{2.1 em}{m}} \\
        }\]
        
        \caption{\textbf{Multivariate Quantum to Quantum Bernoulli factory general circuit.} The circuit takes in input $n_k$ qubits in the state $\ket{z_k}$ for a total of $n_t$ and $m$ ancillary qubits in the state $\ket{0}$.
        The rest of the circuit is similar to that of Fig.~\ref{fig: BF_circuit}.
        As before, this circuit is the most general one implementing a Multivariate Quantum to Quantum Bernoulli factory.}
        \label{fig: MVBF_circuit}
\end{figure}
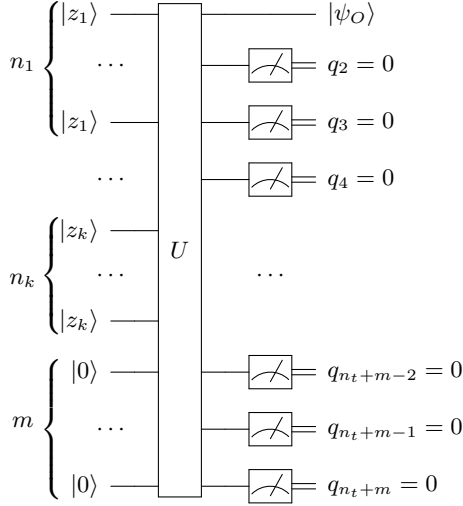

In the following, we present an extension of the Quantum-to-quantum Bernoulli Factory problem, taking into account the possibility of having more than one ``hidden'' state at the input of the circuit. This means that instead of having a state $\ket{z}$, we have a set of different states $\{\ket{z_1}, \dots \ket{z_k}\}$. In addition, the function becomes a multivariate one: $g(z_1, \dots z_k)$. We refer to this new variant of the problem as \emph{Multivariate Quantum-to-quantum Bernoulli factory}. Here, we characterise all relevant aspects of this new extension following the same method previously presented for the original problem.

The more general circuit implementing the Multivariate Quantum-to-quantum Bernoulli factory is the same as the previous case, with the difference that the input can be composed of more than one hidden state, as depicted in Fig.~\ref{fig: MVBF_circuit}.
Analogously to the approach of Sec.~\ref{sec:QQBF}, it can be shown that the set of constructible functions $g(z_1, \dots z_k)$ is a subset of the set of multivariate complex rational functions $\text{Rat}(z_1, \dots, z_k)$.
Moreover, adapting the previous construction method, it is possible to show that for every function of the form $g(z_1, \dots z_k) \in \text{Rat}(z_1, \dots, z_k)$ it is possible to construct a quantum circuit implementing the associated Multivariate Bernoulli factory.
In addition, one can make the circuit use the minimum number of qubits for each variable that corresponds to the lower bound $n_j = \deg_j(g(z_1, \dots z_k))$ and without the necessity of ancillary.  The only exception in the possibility of not using ancillary qubits is when $ \forall j \; \deg_j(g(z_1, \dots z_k)) = 1$, in such a situation an ancillary qubit is needed.
An explicit demonstration is reported in the Supplementary Materials IV.

\section{Multifunctional Quantum-to-quantum Bernoulli factory}

This second extension is motivated by the intrinsic probabilistic nature of the Quantum-to-Quantum Bernoulli Factory, which requires heralding on specific measurement outcomes to implement the desired function. Extending such a consideration, we can say that the Bernoulli factory implements a different function for each measurement outcome. 
Therefore, this leads to the definition of a second extension of the Quantum-to-quantum Bernoulli factory problem, where we examine the possibility of implementing more than one function at the same time.

\begin{figure}[ht!]
        \centering
        \[\Qcircuit @C=1em @R=1em {
        && \lstick{\ket{z}} & \qw & \multigate{6}{U} & \qw & \qw &\rstick{\ket{\psi_O = g_{0/1}(z)}} \qw\\
         && \lstick{\ket{z}} & \qw & \ghost{U} & \qw & \meter & \rstick{q_2 = 0/1} \cw\\
        && \cdots& & \nghost{U} & & \cdots \\
        && \lstick{\ket{z}} & \qw & \ghost{U} & \qw & \meter & \rstick{q_n = 0} \cw\\
        && \lstick{\ket{0_c}} & \qw & \ghost{U} & \qw & \meter & \rstick{q_{n+1} = 0} \cw\\
        && \cdots& & \nghost{U} & & \cdots \\
        && \lstick{\ket{0_c}} & \qw & \ghost{U} & \qw & \meter & \rstick{q_{n+m} = 0} \cw\\
        {\inputgroupv{1}{4}{1em}{3.2 em}{n}} \\
        {\inputgroupv{5}{7}{1em}{2.1 em}{m}} \\
        }\]
        
        \caption{\textbf{Multifunctional Quantum Bernoulli factory general circuit.} This circuit implementing a Multifunctional Quantum Bernoulli factory is similar to the Fig.~\ref{fig: BF_circuit}. The difference is at the measurement stage. Previously, all the measures had to return the outcome $0$. In this case, if the measure $q_2$ returns $0$, the circuit implements the function $g_0(z)$, while if the outcome is 1 $1$ the circuit implements the function $g_1(z)$.
         As before, this circuit is the most general one implementing a Multifunctional Quantum to Quantum Bernoulli factory.
        }
        \label{fig: MFBF_circuit}
\end{figure}
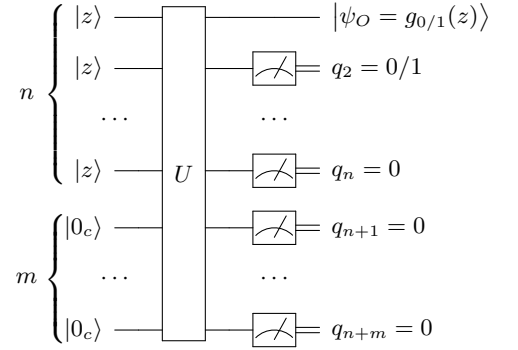

We define a \emph{Multifunctional Quantum-to-quantum Bernoulli factory} associated with a tuple of functions $(g_0, \dots g_l)$ as a QQBF implementing the functions $g_0, \dots g_l$ depending on the output of the heralding measurement. Below, we consider only the simplest case of two functions, but the analysis can also be extended to cases involving more functions.
Starting from considerations similar to those used for the original problem, namely that all the hidden information is only in the input states, we can say that the most general quantum circuit implementing a Multifunctional Quantum-to-quantum Bernoulli factory for two functions is the one depicted in Fig.~\ref{fig: MFBF_circuit}. 
The circuit is identical to the one implementing the QQBF with the difference that the measurement of the second qubit $q_2$ indicates which function $g_0(z)$ or $g_1(z)$ is implemented by the circuit.
Given two simulable functions, it is always possible to construct a Multifunctional Quantum-to-quantum Bernoulli factory (see Supplementary Materials V). However, the success probabilities of individual functions will almost always turn out to be lower than the corresponding probabilities if one constructs a set of Quantum-to-Quantum Bernoulli factories implementing the single functions.

With this consideration in mind, we propose a slightly different definition. More specifically, we introduce a hierarchy for the functions, which means that we prioritise the success probability of some functions over the others. In this context, we say that a function $g_1(z)$ is \emph{compatible} with a function $g_0(z)$ if it is possible to implement a Multifunctional Quantum-to-quantum Bernoulli factory where the success probability of the function $g_0(z)$ is the same achievable with a dedicated Quantum-to-quantum Bernoulli factory.
To understand why we suggest this notion of compatibility, let us consider the following scenario. We suppose that, to implement a specific algorithm, it is necessary to perform two different QQBFs with the same input states that respectively implement the functions $g_0(z)$ and $g_1(z)$. If the function $g_1(z)$ is compatible with the function $g_0(z)$, we can implement a Multivariate Quantum-to-quantum Bernoulli factory without changing the success probability of the function $g_0(z)$, but with the additional functionality that, if the circuit returns a sample from the function $g_1(z)$, we can in principle store the output state to be used in a second moment when needed, increasing the overall efficiency of the algorithm.

Defining $g_0(z) = P(z)/Q(z)$ and $g_1(z) = R(z)/S(z)$, the condition of compatibility of $g_1(z)$ respect $g_0(z)$ is equivalent to the following conditions:
\begin{eqnarray}
    s_1 &=& \sum_j \biggr(\frac{p_j^*}{x}- \frac{q_j^*}{y}\biggl) \frac{r_j}{\binom{n}{j}} = 0, \label{eq:conditions_existance_multi_1} \\
    s_2 &=& \sum_j \biggr(\frac{p_j^*}{x}- \frac{q_j^*}{y}\biggl) \frac{s_j}{\binom{n}{j}} = 0,
    \label{eq:conditions_existance_multi_2}
\end{eqnarray}
where $p_j$, $q_j$, $r_j$, $s_j$ are the coefficients respectively of the polynomials $P(z)$, $Q(z)$, $R(z)$, $S(z)$ and $x$ and $y$ are the values defined by Eqs. \eqref{eq: x_y_variable_1}-\eqref{eq: x_y_variable_2} with $w = 0$.
To prove the statement, we show that the two conditions are necessary and sufficient for the existence of a unitary transformation according to the general circuit of Fig.~\ref{fig: MFBF_circuit}.

Similar to the procedure used in Eq. \eqref{eq: Out_base_2}, we call the first four rows of the unitary transformation as the covectors $\bra{v_0}$, $\bra{v_1}$, $\bra{v_2}$ and $\bra{v_3}$. The first two rows are fixed by the fact that we want to obtain a success probability for $g_{0}(z)$ equal to the one achievable with a dedicated Quantum-to-quantum Bernoulli factory. Hence, $\vert v_{0} \rangle$ and $\vert v_{1} \rangle$ have the following forms:
\begin{equation}
    \ket{v_0} = \sum_{j = 0}^{n} K \frac{p^*_j}{\sqrt{\binom{n}{j}}} \ket{s_j^n} + K x \ket{\theta_0},
\end{equation}
and
\begin{equation}
    \ket{v_1} = \sum_{j = 0}^{n} K \frac{q^*_j}{\sqrt{\binom{n}{j}}} \ket{s_j^n} + K y \ket{\theta_0},
\end{equation}
where $x$, $y$ and $K$ are those found in Eqs.~\eqref{eq: x_y_variable_1}-\eqref{eq: x_y_variable_3} with $w = 0$.
The other two vectors can be expressed in the general forms as:
\begin{equation}
    \ket{v_2} = \sum_{j = 0}^{n} H \frac{r^*_j}{\sqrt{\binom{n}{j}}} \ket{s_j^n} + H a_1 \ket{\theta_0} + H a_2 \ket{\theta_1},
\end{equation}
and
\begin{equation}
    \begin{split}
    \ket{v_3} =& \sum_{j = 0}^{n} H \frac{s^*_j}{\sqrt{\binom{n}{j}}} \ket{s_j^n} + H a_3 \ket{\theta_0} + H a_4 \ket{\theta_1} + \\ &+H a_5 \ket{\theta_2},
    \end{split}
\end{equation}
where $H$ is the normalization factor and $a_1$, $a_2$, $a_3$, $a_4$ and $a_5$ are free parameters.
To ensure that the four vectors represent four rows of a unitary transformation they must be orthonormal between each other as $\langle v_i \vert v_j \rangle = \delta_{i,j}$. 
While some orthonormality conditions are already satistified (namely $\langle v_0 \vert v_0 \rangle = 1$, $\langle v_1 \vert v_1 \rangle = 1$ and $\langle v_0 \vert v_1 \rangle = 0$) the remaining four are the critical ones since there are four equations and two free parameters.
The following two conditions correspond to:
\begin{eqnarray}
 \braket{v_0}{v_2} = 0 \:  &\Rightarrow& \: \sum_j \frac{p_jr_j^*}{\binom{n}{j}} + x^* a_1 = 0,  \\
 \braket{v_1}{v_2} = 0 \: &\Rightarrow& \: \sum_j \frac{q_jr_j^*}{\binom{n}{j}} + y^* a_1 = 0.
\end{eqnarray}
A parameter $a_1$ that solves both the equations exists if Eqs.~\eqref{eq:conditions_existance_multi_1} are satisfied. The other constraint descends from the remaining set of orthonormality conditions.
If Eqs.~\eqref{eq:conditions_existance_multi_1}-\eqref{eq:conditions_existance_multi_2} are satisfied then it is possible to derive the values of $a_1$ and $a_3$. The orthonormality of states $\vert v_{i} \rangle$ allow us always to derive the parameters $a_2$, $a_4$ and $H$. Furthermore fixing the free parameter $a_5 = 0$ we can reach the maximum achievable probability. With this, we have the first four rows of the unitary matrix $U$. By completing to an orthonormal basis we can construct the circuit implementing the desired Multifunctional Quantum-to-quantum Bernoulli factory.

In this way, we prove the necessity and sufficiency of the conditions in Eqs.~\eqref{eq:conditions_existance_multi_1}-\eqref{eq:conditions_existance_multi_2} to test if a function $g_1(z)$ is compatible with a function $g_0(z)$. Also, we explicitly construct a quantum circuit that implements the most efficient Multifunctional Quantum-to-quantum Bernoulli factory associated with the functions.

\section{Application to the sum and product operation}
This section shows an application of the theory developed in this paper for a specific case. In particular, we will construct the multivariate QQBFs corresponding to the sum ($z_1+z_2$) and product ($z_1z_2$) functions and compare them with those presented in the article by Jiang et. al. \cite{Jiang2018} where the QQBF was first proposed.
We will then analyse the compatibility of the two functions and eventually construct the corresponding multifunctional QQBF.

\subsection{Product $z_1z_2$}
The first operation that we analyse is the product operation $g(z_1,z_2) = z_1z_2$ since it has degree one in both variables this tells us that we need only one qubit per variable.
The coefficients of the function are $p_{1,1}=1$, $q_{0,0}=1$, $p_{i,j}=0$ for $(i,j) \neq 1,1$, and $q_{i,j}=0$ for $(i,j) \neq (0,0)$. This gives us $a = 1$, $b = 1$, $c = 0$, leading to $x = 0$, $y = 0$, and $K =1$.
Since $a=b$ and $c=0$, this means that ancillary qubits are not required. The two vectors $v_0$ and $v_1$ are:
\begin{eqnarray}
    \ket{v_0} &=& \ket{00}, \\ 
    \ket{v_1} &=& \ket{11},
\end{eqnarray}
and the final unitary transformation is
\begin{equation}
    U = \ketbra{00}{00}+\ketbra{10}{11}+\ketbra{11}{10}+\ketbra{01}{01},
\end{equation}
which is a CNOT where the first qubit acts as a control and the second qubit is the target, as presented in \cite{Jiang2018}. The corresponding success probability is
\begin{equation}
    \mathrm{Pr}_{\cdot}(z_1,z_2) = \frac{\abs{z_1z_2}^2+1}{(1+\abs{z_1}^2)(1+\abs{z_2}^2)}.
    \label{eq:prod_prob}
\end{equation}

\subsection{Sum  $z_1+z_2$}
The second operation that we analyse is the sum $g(z_1,z_2) = z_1+z_2$. It is also a function of degree one in both variables, which implies that it can be implemented with one qubit for each variable.
The coefficients of the function are $p_{i,i} = 0$, $p_{i,j} = 1$ for $i \neq j$, $q_{0,0}=1$, and $q_{i,j}=0$ for $(i,j) \neq (0,0)$.
This gives us $a = 1$, $b = 2$, $c = 0$, leading to $x = 0$, $y = 1$ and $K =2^{-1/2}$. Since $a\neq b$, this means that the operation requires an ancillary qubit. Choosing $\ket{\theta_0} = \ket{001}$, the two vectors $v_0$ and $v_1$ are:
\begin{eqnarray}
    \ket{v_0} &=& \frac{\ket{010}+\ket{100}}{\sqrt{2}}, \\ 
    \ket{v_1} &=& \frac{\ket{110}+\ket{001}}{\sqrt{2}},
\end{eqnarray}
and the corresponding unitary transformation deduced with the algorithm presented in the Supplementary Materials II is 
\begin{equation}
    U = \begin{pmatrix}
        0 & \frac{1}{\sqrt{2}} & \frac{1}{\sqrt{2}} & 0 & 0 & 0 & 0 & 0 \\
        0 & 0 & 0 & \frac{1}{\sqrt{2}} & \frac{1}{\sqrt{2}} & 0 & 0 & 0 \\
        - \frac{1}{\sqrt{2}} & -\frac{1}{2} & \frac{1}{2} & 0 & 0 & 0& 0 & 0 \\
        \frac{1}{2} & -\frac{1}{2\sqrt{2}} & \frac{1}{2\sqrt{2}} & \frac{1}{2} & -\frac{1}{2} & 0 & 0 & 0\\
        \frac{1}{2} & -\frac{1}{2\sqrt{2}} & \frac{1}{2\sqrt{2}} & -\frac{1}{2} & \frac{1}{2} & 0 & 0 & 0\\
        0 & 0 & 0 & 0 & 0 & 1 & 0 & 0\\
        0 & 0 & 0 & 0 & 0 & 0 & 1 & 0\\
        0 & 0 & 0 & 0 & 0 & 0 & 0 & 1
    \end{pmatrix},
\end{equation}
where the states are ordered with the least significant bit on the left. The corresponding success probability is 
\begin{equation}
    \mathrm{Pr}_+(z_1,z_2) = \frac{\abs{z_1+z_2}^2+1}{2(1+\abs{z_1}^2)(1+\abs{z_2}^2)}.
    \label{eq:sum_prob}
\end{equation}

In Ref. \cite{Jiang2018}, the authors do not propose the sum operation but a slightly different operation where the sum is divided by $\sqrt{2}$, $g(z_1, z_2) = \frac{z_1+z_2}{\sqrt{2}}$. In this case, we find with our construction $a = 2$, $b = 2$, $c = 0$, leading to $x = 0$, $y = 0$, $K =2^{-1/2}$, which implies that there is no need for an ancillary qubit. 
The corresponding unitary transformation is
\begin{equation}
    U = \begin{pmatrix}
        0 & \frac{1}{\sqrt{2}} & \frac{1}{\sqrt{2}} & 0  \\
        0 & 0 & 0 & 1 \\
        1 & 0 & 0 & 0 \\
        0 & \frac{1}{\sqrt{2}} & -\frac{1}{\sqrt{2}} & 0  \\
    \end{pmatrix},
\end{equation}
which is the same unitary matrix found in \cite{Jiang2018}.
The corresponding probability is 
\begin{equation}
    \mathrm{Pr}_+(z_1,z_2) = \frac{\abs{z_1+z_2}^2+2}{2(1+\abs{z_1}^2)(1+\abs{z_2}^2)}.
\end{equation}

As a note, we can construct the circuit for a sum operation by concatenating the previous circuit with a product operation having as input a qubit with $z = \sqrt{2}$. 
However, the total  success probability will be
\begin{equation}
    \mathrm{Pr}_+(z_1,z_2) = \frac{\abs{z_1+z_2}^2+1}{3(1+\abs{z_1}^2)(1+\abs{z_2}^2)}.
\end{equation}
which is lower than the optimal circuit found previously.

\subsection{Compatibility between the sum and the product operation}
In this section, we identify a multifunctional QQBF that implements both the product and the sum operation, testing if the sum operation is compatible with the product one or vice versa.

First, we test if the sum is compatible with the product operation by evaluating the two conditions in Eqs.~\eqref{eq:conditions_existance_multi_1}-\eqref{eq:conditions_existance_multi_2}. Since both $x$ and $y$ are null, we need to modify the functions as $g_0(z_1,z_2) = z_1 z_2+\epsilon$ and then take the limit for $\epsilon \rightarrow 0$. Performing all the calculations we obtain $x = 2^{-1/2} \epsilon$, $y = (3/2)^{1/2}\epsilon$, leading to $s_1 = 0$, $s_2 = (3/2)^{1/2} - 2^{1/2}/\epsilon$. In the limit $\epsilon \rightarrow 0$ the condition $s_2$ is not satisfied, implying that the sum operation is not compatible with the product.

Now we verify whether the product operation is compatible with the sum. Similar to the previous case, since $x = 0$ we need to modify the sum operation $g_0(z_1,z_2) = z_1+z_2+\epsilon$. Performing all calculations described above, we obtain $x = \epsilon(1-\epsilon^2)$, $y = 1+\epsilon^2$, leading to $s_1 = 0$, $s_2 = 2\epsilon^2$. In this case, both conditions are satisfied in the limit of $\epsilon \rightarrow 0$ meaning that the product operation is compatible with the sum operation.
Proceeding with the calculation to find the unitary matrix as described above, we obtain the four vectors
\begin{eqnarray}
    \ket{v_0} &=& \frac{\ket{010}+\ket{100}}{\sqrt{2}}, \\ 
    \ket{v_1} &=& \frac{\ket{110}+\ket{001}}{\sqrt{2}}, \\
    \ket{v_2} &=& \frac{\ket{000}+\ket{101}}{\sqrt{2}}, \\ 
    \ket{v_3} &=& \frac{\ket{110}-\ket{001}}{\sqrt{2}},
\end{eqnarray}
and the final unitary matrix
\begin{equation}
    U = \begin{pmatrix}
        0 & \frac{1}{\sqrt{2}} & \frac{1}{\sqrt{2}} & 0 & 0 & 0 & 0 & 0 \\
        0 & 0 & 0 & \frac{1}{\sqrt{2}} & \frac{1}{\sqrt{2}} & 0 & 0 & 0 \\
        \frac{1}{\sqrt{2}} & 0 & 0 & 0 & 0 & \frac{1}{\sqrt{2}}& 0 & 0 \\
        0 & 0 & 0 & \frac{1}{\sqrt{2}} & -\frac{1}{\sqrt{2}} & 0 & 0 & 0\\
        0 & \frac{1}{\sqrt{2}} & -\frac{1}{\sqrt{2}} & 0 & 0 & 0 & 0 & 0\\
        \frac{1}{\sqrt{2}} & 0 & 0 & 0 & 0 & -\frac{1}{\sqrt{2}} & 0 & 0\\
        0 & 0 & 0 & 0 & 0 & 0 & 1 & 0\\
        0 & 0 & 0 & 0 & 0 & 0 & 0 & 1
    \end{pmatrix}.
\end{equation}
In this case, the success probability of the sum operation is equal to the optimal one of Eq.~\eqref{eq:sum_prob}. On the contrary, the success probability of the product is
\begin{equation}
    Pr_.(z_1,z_2) = \frac{\abs{z_1z_2}^2+1}{2(1+\abs{z_1}^2)(1+\abs{z_2}^2)},
\end{equation}
which is half of the optimal one reported in Eq.~\eqref{eq:prod_prob}.

\section{Discussion}

In this work, we provide a comprehensive analysis the problem of a Quantum-to-Quantum Bernoulli factory. 
The manuscript presents a novel constructive proof for the characterisation of the Quantum-to-Quantum Bernoulli factory, independently establishing the set of simulable functions as the rational functions. While the proof aligns with previous results \cite{Jiang2018}, our approach enables a direct complexity analysis, explicitly quantifying the protocols' qubit requirements and success probability. In particular, we show that the minimum number of required qubits coincides with the degree of the function without the necessity of ancillary qubits, except for degree one, where at most one ancillary qubit is required. We also provide an analytical expression for the maximal success probability and a procedure for the construction of the quantum circuit that saturates those bounds.

Moreover, we formalise and analyse two extensions of the original problem called respectively the multivariate Quantum-to-Quantum Bernoulli factory and the multifunctional Quantum-to-Quantum Bernoulli factory. In those variants we allow, respectively, the possibility to employ more than one "hidden" variable at the input of the circuit or the implementation of more than one function within the same protocol. Similarly to the original problem, we analyse the computational complexity of the two extensions and provide a procedure for the construction of the quantum circuit implementing the desidered protocol.

With those analyses, we provide a comprehensive overview of the protocol, opening the possibility of employing the Quantum-to-Quantum Bernoulli factory as a subroutine for larger quantum computation and quantum communication protocols, and the possibility of more efficient experimental implementation. Furthermore, this work can be at the basis of further theoretical developments of the algorithm and of the proposal of new extensions of the problem, such as different input/output states or different resources employed, such as the one proposed in \cite{Theodore}.



\section*{Acknowledgements}
We acknowledge support from the ERC Advanced Grant QU-BOSS (QUantum advantage via nonlinear BOSon Sampling, grant agreement no. 884676), from the ERC Proof of Concept grant POQUB (Grant Agreement No. 101212872), from the project QU-DICE, Fare Ricerca in Italia, Ministero dell'istruzione e del merito (code: R20TRHTSPA), and from PNRR MUR project PE0000023-NQSTI (National Quantum Science and Technology Institute, Spoke 4).


\begin{thebibliography}{32}%
\makeatletter
\providecommand \@ifxundefined [1]{%
 \@ifx{#1\undefined}
}%
\providecommand \@ifnum [1]{%
 \ifnum #1\expandafter \@firstoftwo
 \else \expandafter \@secondoftwo
 \fi
}%
\providecommand \@ifx [1]{%
 \ifx #1\expandafter \@firstoftwo
 \else \expandafter \@secondoftwo
 \fi
}%
\providecommand \natexlab [1]{#1}%
\providecommand \enquote  [1]{``#1''}%
\providecommand \bibnamefont  [1]{#1}%
\providecommand \bibfnamefont [1]{#1}%
\providecommand \citenamefont [1]{#1}%
\providecommand \href@noop [0]{\@secondoftwo}%
\providecommand \href [0]{\begingroup \@sanitize@url \@href}%
\providecommand \@href[1]{\@@startlink{#1}\@@href}%
\providecommand \@@href[1]{\endgroup#1\@@endlink}%
\providecommand \@sanitize@url [0]{\catcode `\\12\catcode `\$12\catcode
  `\&12\catcode `\#12\catcode `\^12\catcode `\_12\catcode `\%12\relax}%
\providecommand \@@startlink[1]{}%
\providecommand \@@endlink[0]{}%
\providecommand \url  [0]{\begingroup\@sanitize@url \@url }%
\providecommand \@url [1]{\endgroup\@href {#1}{\urlprefix }}%
\providecommand \urlprefix  [0]{URL }%
\providecommand \Eprint [0]{\href }%
\providecommand \doibase [0]{https://doi.org/}%
\providecommand \selectlanguage [0]{\@gobble}%
\providecommand \bibinfo  [0]{\@secondoftwo}%
\providecommand \bibfield  [0]{\@secondoftwo}%
\providecommand \translation [1]{[#1]}%
\providecommand \BibitemOpen [0]{}%
\providecommand \bibitemStop [0]{}%
\providecommand \bibitemNoStop [0]{.\EOS\space}%
\providecommand \EOS [0]{\spacefactor3000\relax}%
\providecommand \BibitemShut  [1]{\csname bibitem#1\endcsname}%
\let\auto@bib@innerbib\@empty
\bibitem [{\citenamefont {Keane}\ and\ \citenamefont
  {O'Brien}(1994)}]{Keane1994}%
  \BibitemOpen
  \bibfield  {author} {\bibinfo {author} {\bibfnamefont {M.~S.}\ \bibnamefont
  {Keane}}\ and\ \bibinfo {author} {\bibfnamefont {G.~L.}\ \bibnamefont
  {O'Brien}},\ }\bibfield  {title} {\bibinfo {title} {A bernoulli factory},\
  }\href {https://doi.org/10.1145/175007.175019} {\bibfield  {journal}
  {\bibinfo  {journal} {ACM Transactions on Modeling and Computer Simulation}\
  }\textbf {\bibinfo {volume} {4}},\ \bibinfo {pages} {213–219} (\bibinfo
  {year} {1994})}\BibitemShut {NoStop}%
\bibitem [{\citenamefont {Paes~Leme}\ and\ \citenamefont
  {Schneider}(2023)}]{PaesLeme2023}%
  \BibitemOpen
  \bibfield  {author} {\bibinfo {author} {\bibfnamefont {R.}~\bibnamefont
  {Paes~Leme}}\ and\ \bibinfo {author} {\bibfnamefont {J.}~\bibnamefont
  {Schneider}},\ }\bibfield  {title} {\bibinfo {title} {Multiparameter
  bernoulli factories},\ }\href {https://doi.org/10.1214/22-aap1913} {\bibfield
   {journal} {\bibinfo  {journal} {The Annals of Applied Probability}\ }\textbf
  {\bibinfo {volume} {33}},\ \bibinfo {pages} {3987} (\bibinfo {year}
  {2023})}\BibitemShut {NoStop}%
\bibitem [{\citenamefont {Niazadeh}\ \emph {et~al.}(2024)\citenamefont
  {Niazadeh}, \citenamefont {Leme},\ and\ \citenamefont
  {Schneider}}]{Niazadeh2024}%
  \BibitemOpen
  \bibfield  {author} {\bibinfo {author} {\bibfnamefont {R.}~\bibnamefont
  {Niazadeh}}, \bibinfo {author} {\bibfnamefont {R.~P.}\ \bibnamefont {Leme}},\
  and\ \bibinfo {author} {\bibfnamefont {J.}~\bibnamefont {Schneider}},\
  }\bibfield  {title} {\bibinfo {title} {Bernoulli factories for flow-based
  polytopes},\ }\href {https://doi.org/10.1137/23m1558343} {\bibfield
  {journal} {\bibinfo  {journal} {SIAM Journal on Discrete Mathematics}\
  }\textbf {\bibinfo {volume} {38}},\ \bibinfo {pages} {726–742} (\bibinfo
  {year} {2024})}\BibitemShut {NoStop}%
\bibitem [{\citenamefont {Niazadeh}\ \emph {et~al.}(2023)\citenamefont
  {Niazadeh}, \citenamefont {Paes~Leme},\ and\ \citenamefont
  {Schneider}}]{Niazadeh2023}%
  \BibitemOpen
  \bibfield  {author} {\bibinfo {author} {\bibfnamefont {R.}~\bibnamefont
  {Niazadeh}}, \bibinfo {author} {\bibfnamefont {R.}~\bibnamefont
  {Paes~Leme}},\ and\ \bibinfo {author} {\bibfnamefont {J.}~\bibnamefont
  {Schneider}},\ }\bibfield  {title} {\bibinfo {title} {Combinatorial bernoulli
  factories},\ }\href {https://doi.org/10.3150/22-bej1497} {\bibfield
  {journal} {\bibinfo  {journal} {Bernoulli}\ }\textbf {\bibinfo {volume}
  {29}},\ \bibinfo {pages} {1246} (\bibinfo {year} {2023})}\BibitemShut
  {NoStop}%
\bibitem [{\citenamefont {Flegal}\ and\ \citenamefont
  {Herbei}(2012)}]{flegal2012exact}%
  \BibitemOpen
  \bibfield  {author} {\bibinfo {author} {\bibfnamefont {J.~M.}\ \bibnamefont
  {Flegal}}\ and\ \bibinfo {author} {\bibfnamefont {R.}~\bibnamefont
  {Herbei}},\ }\bibfield  {title} {\bibinfo {title} {Exact sampling for
  intractable probability distributions via a bernoulli factory},\ }\href
  {https://doi.org/10.1214/11-ejs663} {\bibfield  {journal} {\bibinfo
  {journal} {Electronic Journal of Statistics}\ }\textbf {\bibinfo {volume}
  {6}},\ \bibinfo {pages} {10} (\bibinfo {year} {2012})}\BibitemShut {NoStop}%
\bibitem [{\citenamefont {Vats}\ \emph {et~al.}(2021)\citenamefont {Vats},
  \citenamefont {Gon\c{c}alves}, \citenamefont {Łatuszyński},\ and\
  \citenamefont {Roberts}}]{vats2022efficient}%
  \BibitemOpen
  \bibfield  {author} {\bibinfo {author} {\bibfnamefont {D.}~\bibnamefont
  {Vats}}, \bibinfo {author} {\bibfnamefont {F.~B.}\ \bibnamefont
  {Gon\c{c}alves}}, \bibinfo {author} {\bibfnamefont {K.}~\bibnamefont
  {Łatuszyński}},\ and\ \bibinfo {author} {\bibfnamefont {G.~O.}\
  \bibnamefont {Roberts}},\ }\bibfield  {title} {\bibinfo {title} {Efficient
  bernoulli factory markov chain monte carlo for intractable posteriors},\
  }\href {https://doi.org/10.1093/biomet/asab031} {\bibfield  {journal}
  {\bibinfo  {journal} {Biometrika}\ }\textbf {\bibinfo {volume} {109}},\
  \bibinfo {pages} {369–385} (\bibinfo {year} {2021})}\BibitemShut {NoStop}%
\bibitem [{\citenamefont {Gon\c{c}alves}\ \emph {et~al.}(2023)\citenamefont
  {Gon\c{c}alves}, \citenamefont {Łatuszyński},\ and\ \citenamefont
  {Roberts}}]{gonccalves2023exact}%
  \BibitemOpen
  \bibfield  {author} {\bibinfo {author} {\bibfnamefont {F.~B.}\ \bibnamefont
  {Gon\c{c}alves}}, \bibinfo {author} {\bibfnamefont {K.}~\bibnamefont
  {Łatuszyński}},\ and\ \bibinfo {author} {\bibfnamefont {G.~O.}\
  \bibnamefont {Roberts}},\ }\bibfield  {title} {\bibinfo {title} {Exact monte
  carlo likelihood-based inference for jump-diffusion processes},\ }\href
  {https://doi.org/10.1093/jrsssb/qkad022} {\bibfield  {journal} {\bibinfo
  {journal} {Journal of the Royal Statistical Society Series B: Statistical
  Methodology}\ }\textbf {\bibinfo {volume} {85}},\ \bibinfo {pages}
  {732–756} (\bibinfo {year} {2023})}\BibitemShut {NoStop}%
\bibitem [{\citenamefont {Herbei}\ and\ \citenamefont
  {Berliner}(2014)}]{herbei2014estimating}%
  \BibitemOpen
  \bibfield  {author} {\bibinfo {author} {\bibfnamefont {R.}~\bibnamefont
  {Herbei}}\ and\ \bibinfo {author} {\bibfnamefont {L.~M.}\ \bibnamefont
  {Berliner}},\ }\bibfield  {title} {\bibinfo {title} {Estimating ocean
  circulation: An mcmc approach with approximated likelihoods via the bernoulli
  factory},\ }\href {https://doi.org/10.1080/01621459.2014.914439} {\bibfield
  {journal} {\bibinfo  {journal} {Journal of the American Statistical
  Association}\ }\textbf {\bibinfo {volume} {109}},\ \bibinfo {pages}
  {944–954} (\bibinfo {year} {2014})}\BibitemShut {NoStop}%
\bibitem [{\citenamefont {Koskela}\ \emph {et~al.}(2023)\citenamefont
  {Koskela}, \citenamefont {Łatuszyński},\ and\ \citenamefont
  {Spanò}}]{Koskela2023}%
  \BibitemOpen
  \bibfield  {author} {\bibinfo {author} {\bibfnamefont {J.}~\bibnamefont
  {Koskela}}, \bibinfo {author} {\bibfnamefont {K.}~\bibnamefont
  {Łatuszyński}},\ and\ \bibinfo {author} {\bibfnamefont {D.}~\bibnamefont
  {Spanò}},\ }\href@noop {} {\bibinfo {title} {Bernoulli factories and duality
  in wright-fisher and allen-cahn models of population genetics}} (\bibinfo
  {year} {2023}),\ \Eprint {https://arxiv.org/abs/2306.03539} {arXiv:2306.03539
  [math]} \BibitemShut {NoStop}%
\bibitem [{\citenamefont {Morina}\ \emph {et~al.}(2022)\citenamefont {Morina},
  \citenamefont {Łatuszyński}, \citenamefont {Nayar},\ and\ \citenamefont
  {Wendland}}]{Morina2022}%
  \BibitemOpen
  \bibfield  {author} {\bibinfo {author} {\bibfnamefont {G.}~\bibnamefont
  {Morina}}, \bibinfo {author} {\bibfnamefont {K.}~\bibnamefont
  {Łatuszyński}}, \bibinfo {author} {\bibfnamefont {P.}~\bibnamefont
  {Nayar}},\ and\ \bibinfo {author} {\bibfnamefont {A.}~\bibnamefont
  {Wendland}},\ }\bibfield  {title} {\bibinfo {title} {From the bernoulli
  factory to a dice enterprise via perfect sampling of markov chains},\ }\href
  {https://doi.org/10.1214/21-aap1679} {\bibfield  {journal} {\bibinfo
  {journal} {The Annals of Applied Probability}\ }\textbf {\bibinfo {volume}
  {32}},\ \bibinfo {pages} {327} (\bibinfo {year} {2022})}\BibitemShut
  {NoStop}%
\bibitem [{\citenamefont {Goyal}\ and\ \citenamefont
  {Sigman}(2012)}]{Goyal2012}%
  \BibitemOpen
  \bibfield  {author} {\bibinfo {author} {\bibfnamefont {V.}~\bibnamefont
  {Goyal}}\ and\ \bibinfo {author} {\bibfnamefont {K.}~\bibnamefont {Sigman}},\
  }\bibfield  {title} {\bibinfo {title} {On simulating a class of bernstein
  polynomials},\ }\href {https://doi.org/10.1145/2133390.2133396} {\bibfield
  {journal} {\bibinfo  {journal} {ACM Transactions on Modeling and Computer
  Simulation}\ }\textbf {\bibinfo {volume} {22}},\ \bibinfo {pages} {1–5}
  (\bibinfo {year} {2012})}\BibitemShut {NoStop}%
\bibitem [{\citenamefont {Mossel}\ and\ \citenamefont
  {Peres}(2005)}]{Mossel2005}%
  \BibitemOpen
  \bibfield  {author} {\bibinfo {author} {\bibfnamefont {E.}~\bibnamefont
  {Mossel}}\ and\ \bibinfo {author} {\bibfnamefont {Y.}~\bibnamefont {Peres}},\
  }\bibfield  {title} {\bibinfo {title} {New coins from old: Computing with
  unknown bias},\ }\href {https://doi.org/10.1007/s00493-005-0043-1} {\bibfield
   {journal} {\bibinfo  {journal} {Combinatorica}\ }\textbf {\bibinfo {volume}
  {25}},\ \bibinfo {pages} {707–724} (\bibinfo {year} {2005})}\BibitemShut
  {NoStop}%
\bibitem [{\citenamefont {Łatuszyński}\ \emph {et~al.}(2011)\citenamefont
  {Łatuszyński}, \citenamefont {Kosmidis}, \citenamefont {Papaspiliopoulos},\
  and\ \citenamefont {Roberts}}]{atuszyski2011}%
  \BibitemOpen
  \bibfield  {author} {\bibinfo {author} {\bibfnamefont {K.}~\bibnamefont
  {Łatuszyński}}, \bibinfo {author} {\bibfnamefont {I.}~\bibnamefont
  {Kosmidis}}, \bibinfo {author} {\bibfnamefont {O.}~\bibnamefont
  {Papaspiliopoulos}},\ and\ \bibinfo {author} {\bibfnamefont {G.~O.}\
  \bibnamefont {Roberts}},\ }\bibfield  {title} {\bibinfo {title} {Simulating
  events of unknown probabilities via reverse time martingales},\ }\href
  {https://doi.org/10.1002/rsa.20333} {\bibfield  {journal} {\bibinfo
  {journal} {Random Structures \& Algorithms}\ }\textbf {\bibinfo {volume}
  {38}},\ \bibinfo {pages} {441–452} (\bibinfo {year} {2011})}\BibitemShut
  {NoStop}%
\bibitem [{\citenamefont {Nacu}\ and\ \citenamefont {Peres}(2005)}]{Nacu2005}%
  \BibitemOpen
  \bibfield  {author} {\bibinfo {author} {\bibfnamefont {S.}~\bibnamefont
  {Nacu}}\ and\ \bibinfo {author} {\bibfnamefont {Y.}~\bibnamefont {Peres}},\
  }\bibfield  {title} {\bibinfo {title} {{Fast simulation of new coins from
  old}},\ }\href {https://doi.org/10.1214/105051604000000549} {\bibfield
  {journal} {\bibinfo  {journal} {The Annals of Applied Probability}\ }\textbf
  {\bibinfo {volume} {15}},\ \bibinfo {pages} {93 } (\bibinfo {year}
  {2005})}\BibitemShut {NoStop}%
\bibitem [{\citenamefont {Mendo}(2019)}]{Mendo2019}%
  \BibitemOpen
  \bibfield  {author} {\bibinfo {author} {\bibfnamefont {L.}~\bibnamefont
  {Mendo}},\ }\bibfield  {title} {\bibinfo {title} {An asymptotically optimal
  bernoulli factory for certain functions that can be expressed as power
  series},\ }\href {https://doi.org/10.1016/j.spa.2018.11.017} {\bibfield
  {journal} {\bibinfo  {journal} {Stochastic Processes and their Applications}\
  }\textbf {\bibinfo {volume} {129}},\ \bibinfo {pages} {4366–4384} (\bibinfo
  {year} {2019})}\BibitemShut {NoStop}%
\bibitem [{\citenamefont {Flajolet}\ \emph {et~al.}(2011)\citenamefont
  {Flajolet}, \citenamefont {Pelletier},\ and\ \citenamefont
  {Soria}}]{Flajolet2011}%
  \BibitemOpen
  \bibfield  {author} {\bibinfo {author} {\bibfnamefont {P.}~\bibnamefont
  {Flajolet}}, \bibinfo {author} {\bibfnamefont {M.}~\bibnamefont
  {Pelletier}},\ and\ \bibinfo {author} {\bibfnamefont {M.}~\bibnamefont
  {Soria}},\ }\bibfield  {title} {\bibinfo {title} {On buffon machines and
  numbers},\ }in\ \href {https://doi.org/10.1137/1.9781611973082.15} {\emph
  {\bibinfo {booktitle} {Proceedings of the Twenty-Second Annual ACM-SIAM
  Symposium on Discrete Algorithms}}}\ (\bibinfo  {publisher} {Society for
  Industrial and Applied Mathematics},\ \bibinfo {year} {2011})\BibitemShut
  {NoStop}%
\bibitem [{\citenamefont {Holtz}\ \emph {et~al.}(2010)\citenamefont {Holtz},
  \citenamefont {Nazarov},\ and\ \citenamefont {Peres}}]{Holtz2010}%
  \BibitemOpen
  \bibfield  {author} {\bibinfo {author} {\bibfnamefont {O.}~\bibnamefont
  {Holtz}}, \bibinfo {author} {\bibfnamefont {F.}~\bibnamefont {Nazarov}},\
  and\ \bibinfo {author} {\bibfnamefont {Y.}~\bibnamefont {Peres}},\ }\bibfield
   {title} {\bibinfo {title} {New coins from old, smoothly},\ }\href
  {https://doi.org/10.1007/s00365-010-9108-5} {\bibfield  {journal} {\bibinfo
  {journal} {Constructive Approximation}\ }\textbf {\bibinfo {volume} {33}},\
  \bibinfo {pages} {331–363} (\bibinfo {year} {2010})}\BibitemShut {NoStop}%
\bibitem [{\citenamefont {Dale}\ \emph {et~al.}(2015)\citenamefont {Dale},
  \citenamefont {Jennings},\ and\ \citenamefont {Rudolph}}]{Dale2015}%
  \BibitemOpen
  \bibfield  {author} {\bibinfo {author} {\bibfnamefont {H.}~\bibnamefont
  {Dale}}, \bibinfo {author} {\bibfnamefont {D.}~\bibnamefont {Jennings}},\
  and\ \bibinfo {author} {\bibfnamefont {T.}~\bibnamefont {Rudolph}},\
  }\bibfield  {title} {\bibinfo {title} {Provable quantum advantage in
  randomness processing},\ }\href {https://doi.org/10.1038/ncomms9203}
  {\bibfield  {journal} {\bibinfo  {journal} {Nature Communications}\ }\textbf
  {\bibinfo {volume} {6}} (\bibinfo {year} {2015})}\BibitemShut {NoStop}%
\bibitem [{\citenamefont {Dale}(2016)}]{Dale2016}%
  \BibitemOpen
  \bibfield  {author} {\bibinfo {author} {\bibfnamefont {H.}~\bibnamefont
  {Dale}},\ }\emph {\bibinfo {title} {Quantum coins and quantum sampling}},\
  \href {https://doi.org/10.25560/49203} {Ph.D. thesis},\ \bibinfo  {school}
  {Imperial College London} (\bibinfo {year} {2016})\BibitemShut {NoStop}%
\bibitem [{\citenamefont {Patel}\ \emph {et~al.}(2019)\citenamefont {Patel},
  \citenamefont {Rudolph},\ and\ \citenamefont {Pryde}}]{Patel2019}%
  \BibitemOpen
  \bibfield  {author} {\bibinfo {author} {\bibfnamefont {R.~B.}\ \bibnamefont
  {Patel}}, \bibinfo {author} {\bibfnamefont {T.}~\bibnamefont {Rudolph}},\
  and\ \bibinfo {author} {\bibfnamefont {G.~J.}\ \bibnamefont {Pryde}},\
  }\bibfield  {title} {\bibinfo {title} {An experimental quantum bernoulli
  factory},\ }\href {https://doi.org/10.1126/sciadv.aau6668} {\bibfield
  {journal} {\bibinfo  {journal} {Science Advances}\ }\textbf {\bibinfo
  {volume} {5}},\ \bibinfo {pages} {eaau6668} (\bibinfo {year}
  {2019})}\BibitemShut {NoStop}%
\bibitem [{\citenamefont {Yuan}\ \emph {et~al.}(2016)\citenamefont {Yuan},
  \citenamefont {Liu}, \citenamefont {Xu}, \citenamefont {Wang}, \citenamefont
  {Ma}, \citenamefont {Zhang}, \citenamefont {Yan}, \citenamefont {Vijay},
  \citenamefont {Sun},\ and\ \citenamefont {Ma}}]{yuan2016}%
  \BibitemOpen
  \bibfield  {author} {\bibinfo {author} {\bibfnamefont {X.}~\bibnamefont
  {Yuan}}, \bibinfo {author} {\bibfnamefont {K.}~\bibnamefont {Liu}}, \bibinfo
  {author} {\bibfnamefont {Y.}~\bibnamefont {Xu}}, \bibinfo {author}
  {\bibfnamefont {W.}~\bibnamefont {Wang}}, \bibinfo {author} {\bibfnamefont
  {Y.}~\bibnamefont {Ma}}, \bibinfo {author} {\bibfnamefont {F.}~\bibnamefont
  {Zhang}}, \bibinfo {author} {\bibfnamefont {Z.}~\bibnamefont {Yan}}, \bibinfo
  {author} {\bibfnamefont {R.}~\bibnamefont {Vijay}}, \bibinfo {author}
  {\bibfnamefont {L.}~\bibnamefont {Sun}},\ and\ \bibinfo {author}
  {\bibfnamefont {X.}~\bibnamefont {Ma}},\ }\bibfield  {title} {\bibinfo
  {title} {Experimental quantum randomness processing using superconducting
  qubits},\ }\href {https://doi.org/10.1103/physrevlett.117.010502} {\bibfield
  {journal} {\bibinfo  {journal} {Physical Review Letters}\ }\textbf {\bibinfo
  {volume} {117}},\ \bibinfo {pages} {010502} (\bibinfo {year}
  {2016})}\BibitemShut {NoStop}%
\bibitem [{\citenamefont {Jiang}\ \emph {et~al.}(2018)\citenamefont {Jiang},
  \citenamefont {Zhang},\ and\ \citenamefont {Sun}}]{Jiang2018}%
  \BibitemOpen
  \bibfield  {author} {\bibinfo {author} {\bibfnamefont {J.}~\bibnamefont
  {Jiang}}, \bibinfo {author} {\bibfnamefont {J.}~\bibnamefont {Zhang}},\ and\
  \bibinfo {author} {\bibfnamefont {X.}~\bibnamefont {Sun}},\ }\bibfield
  {title} {\bibinfo {title} {Quantum-to-quantum bernoulli factory problem},\
  }\href {https://doi.org/10.1103/physreva.97.032303} {\bibfield  {journal}
  {\bibinfo  {journal} {Physical Review A}\ }\textbf {\bibinfo {volume} {97}}
  (\bibinfo {year} {2018})}\BibitemShut {NoStop}%
\bibitem [{\citenamefont {Zhan}\ \emph {et~al.}(2020)\citenamefont {Zhan},
  \citenamefont {Wang}, \citenamefont {Xiao}, \citenamefont {Bian},\ and\
  \citenamefont {Xue}}]{Zhan2020}%
  \BibitemOpen
  \bibfield  {author} {\bibinfo {author} {\bibfnamefont {X.}~\bibnamefont
  {Zhan}}, \bibinfo {author} {\bibfnamefont {K.}~\bibnamefont {Wang}}, \bibinfo
  {author} {\bibfnamefont {L.}~\bibnamefont {Xiao}}, \bibinfo {author}
  {\bibfnamefont {Z.}~\bibnamefont {Bian}},\ and\ \bibinfo {author}
  {\bibfnamefont {P.}~\bibnamefont {Xue}},\ }\bibfield  {title} {\bibinfo
  {title} {Experimental demonstration of quantum-to-quantum bernoulli
  factory},\ }\href {https://doi.org/10.1103/physreva.102.012605} {\bibfield
  {journal} {\bibinfo  {journal} {Physical Review A}\ }\textbf {\bibinfo
  {volume} {102}},\ \bibinfo {pages} {012605} (\bibinfo {year}
  {2020})}\BibitemShut {NoStop}%
\bibitem [{\citenamefont {Liu}\ \emph {et~al.}(2021)\citenamefont {Liu},
  \citenamefont {Jiang}, \citenamefont {Zhu}, \citenamefont {Wang},
  \citenamefont {Ding}, \citenamefont {Qiang}, \citenamefont {Huang},
  \citenamefont {Xu}, \citenamefont {Zhang}, \citenamefont {Tian},
  \citenamefont {Fu}, \citenamefont {Deng}, \citenamefont {Wu}, \citenamefont
  {Sun}, \citenamefont {Yang},\ and\ \citenamefont {Wu}}]{liu2021general}%
  \BibitemOpen
  \bibfield  {author} {\bibinfo {author} {\bibfnamefont {Y.}~\bibnamefont
  {Liu}}, \bibinfo {author} {\bibfnamefont {J.}~\bibnamefont {Jiang}}, \bibinfo
  {author} {\bibfnamefont {P.}~\bibnamefont {Zhu}}, \bibinfo {author}
  {\bibfnamefont {D.}~\bibnamefont {Wang}}, \bibinfo {author} {\bibfnamefont
  {J.}~\bibnamefont {Ding}}, \bibinfo {author} {\bibfnamefont {X.}~\bibnamefont
  {Qiang}}, \bibinfo {author} {\bibfnamefont {A.}~\bibnamefont {Huang}},
  \bibinfo {author} {\bibfnamefont {P.}~\bibnamefont {Xu}}, \bibinfo {author}
  {\bibfnamefont {J.}~\bibnamefont {Zhang}}, \bibinfo {author} {\bibfnamefont
  {G.}~\bibnamefont {Tian}}, \bibinfo {author} {\bibfnamefont {X.}~\bibnamefont
  {Fu}}, \bibinfo {author} {\bibfnamefont {M.}~\bibnamefont {Deng}}, \bibinfo
  {author} {\bibfnamefont {C.}~\bibnamefont {Wu}}, \bibinfo {author}
  {\bibfnamefont {X.}~\bibnamefont {Sun}}, \bibinfo {author} {\bibfnamefont
  {X.}~\bibnamefont {Yang}},\ and\ \bibinfo {author} {\bibfnamefont
  {J.}~\bibnamefont {Wu}},\ }\bibfield  {title} {\bibinfo {title} {General
  quantum bernoulli factory: framework analysis and experiments},\ }\href
  {https://doi.org/10.1088/2058-9565/ac2061} {\bibfield  {journal} {\bibinfo
  {journal} {Quantum Science and Technology}\ }\textbf {\bibinfo {volume}
  {6}},\ \bibinfo {pages} {045025} (\bibinfo {year} {2021})}\BibitemShut
  {NoStop}%
\bibitem [{\citenamefont {Hoch}\ \emph {et~al.}(2024)\citenamefont {Hoch},
  \citenamefont {Giordani}, \citenamefont {Castello}, \citenamefont {Carvacho},
  \citenamefont {Spagnolo}, \citenamefont {Ceccarelli}, \citenamefont
  {Pentangelo}, \citenamefont {Piacentini}, \citenamefont {Crespi},
  \citenamefont {Osellame}, \citenamefont {Galvão},\ and\ \citenamefont
  {Sciarrino}}]{Hoch2024}%
  \BibitemOpen
  \bibfield  {author} {\bibinfo {author} {\bibfnamefont {F.}~\bibnamefont
  {Hoch}}, \bibinfo {author} {\bibfnamefont {T.}~\bibnamefont {Giordani}},
  \bibinfo {author} {\bibfnamefont {L.}~\bibnamefont {Castello}}, \bibinfo
  {author} {\bibfnamefont {G.}~\bibnamefont {Carvacho}}, \bibinfo {author}
  {\bibfnamefont {N.}~\bibnamefont {Spagnolo}}, \bibinfo {author}
  {\bibfnamefont {F.}~\bibnamefont {Ceccarelli}}, \bibinfo {author}
  {\bibfnamefont {C.}~\bibnamefont {Pentangelo}}, \bibinfo {author}
  {\bibfnamefont {S.}~\bibnamefont {Piacentini}}, \bibinfo {author}
  {\bibfnamefont {A.}~\bibnamefont {Crespi}}, \bibinfo {author} {\bibfnamefont
  {R.}~\bibnamefont {Osellame}}, \bibinfo {author} {\bibfnamefont {E.~F.}\
  \bibnamefont {Galvão}},\ and\ \bibinfo {author} {\bibfnamefont
  {F.}~\bibnamefont {Sciarrino}},\ }\bibfield  {title} {\bibinfo {title}
  {Modular quantum-to-quantum bernoulli factory in an integrated photonic
  processor},\ }\href {https://doi.org/10.1038/s41566-024-01526-8} {\bibfield
  {journal} {\bibinfo  {journal} {Nature Photonics}\ }\textbf {\bibinfo
  {volume} {19}},\ \bibinfo {pages} {12–19} (\bibinfo {year}
  {2024})}\BibitemShut {NoStop}%
\bibitem [{\citenamefont {Rodari}\ \emph {et~al.}(2024)\citenamefont {Rodari},
  \citenamefont {Hoch}, \citenamefont {Suprano}, \citenamefont {Giordani},
  \citenamefont {Negro}, \citenamefont {Carvacho}, \citenamefont {Spagnolo},
  \citenamefont {Galvão},\ and\ \citenamefont {Sciarrino}}]{Rodari2024}%
  \BibitemOpen
  \bibfield  {author} {\bibinfo {author} {\bibfnamefont {G.}~\bibnamefont
  {Rodari}}, \bibinfo {author} {\bibfnamefont {F.}~\bibnamefont {Hoch}},
  \bibinfo {author} {\bibfnamefont {A.}~\bibnamefont {Suprano}}, \bibinfo
  {author} {\bibfnamefont {T.}~\bibnamefont {Giordani}}, \bibinfo {author}
  {\bibfnamefont {E.}~\bibnamefont {Negro}}, \bibinfo {author} {\bibfnamefont
  {G.}~\bibnamefont {Carvacho}}, \bibinfo {author} {\bibfnamefont
  {N.}~\bibnamefont {Spagnolo}}, \bibinfo {author} {\bibfnamefont {E.~F.}\
  \bibnamefont {Galvão}},\ and\ \bibinfo {author} {\bibfnamefont
  {F.}~\bibnamefont {Sciarrino}},\ }\bibfield  {title} {\bibinfo {title}
  {Polarization-encoded photonic quantum-to-quantum bernoulli factory based on
  a quantum dot source},\ }\href {https://doi.org/10.1126/sciadv.ado6244}
  {\bibfield  {journal} {\bibinfo  {journal} {Science Advances}\ }\textbf
  {\bibinfo {volume} {10}},\ \bibinfo {pages} {eado6244} (\bibinfo {year}
  {2024})}\BibitemShut {NoStop}%
\bibitem [{\citenamefont {Kashefi}\ and\ \citenamefont
  {Pappa}(2017)}]{Kashefi2017}%
  \BibitemOpen
  \bibfield  {author} {\bibinfo {author} {\bibfnamefont {E.}~\bibnamefont
  {Kashefi}}\ and\ \bibinfo {author} {\bibfnamefont {A.}~\bibnamefont
  {Pappa}},\ }\bibfield  {title} {\bibinfo {title} {Multiparty delegated
  quantum computing},\ }\href {https://doi.org/10.3390/cryptography1020012}
  {\bibfield  {journal} {\bibinfo  {journal} {Cryptography}\ }\textbf {\bibinfo
  {volume} {1}},\ \bibinfo {pages} {12} (\bibinfo {year} {2017})}\BibitemShut
  {NoStop}%
\bibitem [{\citenamefont {Fitzsimons}\ and\ \citenamefont
  {Kashefi}(2017)}]{Fitzsimons2017}%
  \BibitemOpen
  \bibfield  {author} {\bibinfo {author} {\bibfnamefont {J.~F.}\ \bibnamefont
  {Fitzsimons}}\ and\ \bibinfo {author} {\bibfnamefont {E.}~\bibnamefont
  {Kashefi}},\ }\bibfield  {title} {\bibinfo {title} {Unconditionally
  verifiable blind quantum computation},\ }\href
  {https://doi.org/10.1103/physreva.96.012303} {\bibfield  {journal} {\bibinfo
  {journal} {Physical Review A}\ }\textbf {\bibinfo {volume} {96}},\ \bibinfo
  {pages} {012303} (\bibinfo {year} {2017})}\BibitemShut {NoStop}%
\bibitem [{\citenamefont {Polacchi}\ \emph {et~al.}(2023)\citenamefont
  {Polacchi}, \citenamefont {Leichtle}, \citenamefont {Limongi}, \citenamefont
  {Carvacho}, \citenamefont {Milani}, \citenamefont {Spagnolo}, \citenamefont
  {Kaplan}, \citenamefont {Sciarrino},\ and\ \citenamefont
  {Kashefi}}]{Polacchi2023}%
  \BibitemOpen
  \bibfield  {author} {\bibinfo {author} {\bibfnamefont {B.}~\bibnamefont
  {Polacchi}}, \bibinfo {author} {\bibfnamefont {D.}~\bibnamefont {Leichtle}},
  \bibinfo {author} {\bibfnamefont {L.}~\bibnamefont {Limongi}}, \bibinfo
  {author} {\bibfnamefont {G.}~\bibnamefont {Carvacho}}, \bibinfo {author}
  {\bibfnamefont {G.}~\bibnamefont {Milani}}, \bibinfo {author} {\bibfnamefont
  {N.}~\bibnamefont {Spagnolo}}, \bibinfo {author} {\bibfnamefont
  {M.}~\bibnamefont {Kaplan}}, \bibinfo {author} {\bibfnamefont
  {F.}~\bibnamefont {Sciarrino}},\ and\ \bibinfo {author} {\bibfnamefont
  {E.}~\bibnamefont {Kashefi}},\ }\bibfield  {title} {\bibinfo {title}
  {Multi-client distributed blind quantum computation with the qline
  architecture},\ }\href {https://doi.org/10.1038/s41467-023-43617-0}
  {\bibfield  {journal} {\bibinfo  {journal} {Nature Communications}\ }\textbf
  {\bibinfo {volume} {14}},\ \bibinfo {pages} {7743} (\bibinfo {year}
  {2023})}\BibitemShut {NoStop}%
\bibitem [{\citenamefont {Nielsen}\ and\ \citenamefont
  {Chuang}(2009)}]{Nielsen2009}%
  \BibitemOpen
  \bibfield  {author} {\bibinfo {author} {\bibfnamefont {M.~A.}\ \bibnamefont
  {Nielsen}}\ and\ \bibinfo {author} {\bibfnamefont {I.~L.}\ \bibnamefont
  {Chuang}},\ }\href {https://doi.org/10.1017/cbo9780511976667} {\emph
  {\bibinfo {title} {Quantum Computation and Quantum Information}}}\ (\bibinfo
  {publisher} {Cambridge University Press},\ \bibinfo {year}
  {2009})\BibitemShut {NoStop}%
\bibitem [{\citenamefont {Gurevich}\ and\ \citenamefont
  {Blass}(2021)}]{Gurevich2021}%
  \BibitemOpen
  \bibfield  {author} {\bibinfo {author} {\bibfnamefont {Y.}~\bibnamefont
  {Gurevich}}\ and\ \bibinfo {author} {\bibfnamefont {A.}~\bibnamefont
  {Blass}},\ }\href@noop {} {\bibinfo {title} {Quantum circuits with classical
  channels and the principle of deferred measurements}} (\bibinfo {year}
  {2021}),\ \Eprint {https://arxiv.org/abs/2107.08324} {arXiv:2107.08324
  [quant-ph]} \BibitemShut {NoStop}%
\bibitem [{\citenamefont {Yoder}(2015)}]{Theodore}%
  \BibitemOpen
  \bibfield  {author} {\bibinfo {author} {\bibfnamefont {T.~J.}\ \bibnamefont
  {Yoder}},\ }\href {https://www.scottaaronson.com/6s899/tedyoder.pdf}
  {\bibinfo {title} {Building and bounding quantum bernoulli factories}}
  (\bibinfo {year} {2015})\BibitemShut {NoStop}%
\end{thebibliography}

%

\end{document}


\title{Supplementary materials: Complexity and multi-functional variants of Quantum-to-Quantum Bernoulli Factories}

\author{Francesco Hoch} 
\affiliation{Dipartimento di Fisica, Sapienza Universit\`{a} di Roma,
Piazzale Aldo Moro 5, I-00185 Roma, Italy}

\author{Taira Giordani} 
\affiliation{Dipartimento di Fisica, Sapienza Universit\`{a} di Roma,
Piazzale Aldo Moro 5, I-00185 Roma, Italy}

\author{Gonzalo Carvacho}
\affiliation{Dipartimento di Fisica, Sapienza Universit\`{a} di Roma,
Piazzale Aldo Moro 5, I-00185 Roma, Italy}

\author{Nicol\`o Spagnolo}
\affiliation{Dipartimento di Fisica, Sapienza Universit\`{a} di Roma,
Piazzale Aldo Moro 5, I-00185 Roma, Italy}

\author{Fabio Sciarrino}
\email[Corresponding author: ]{fabio.sciarrino@uniroma1.it}
\affiliation{Dipartimento di Fisica, Sapienza Universit\`{a} di Roma,
Piazzale Aldo Moro 5, I-00185 Roma, Italy}

\maketitle

\section{$\ket{v_0}$ and $\ket{v_1}$ general form}
In this Section, we present an in-depth demonstration of the general form for the vectors $\ket{v_0}$ and $\ket{v_1}$ from the conditions given in the main text, which we recall here:
\begin{eqnarray}
    \braket{v_0}{s_j^n} &=&K \frac{p_j}{\sqrt{\binom{n}{j}}}, \\  
    \braket{v_1}{s_j^n} &=&K \frac{q_j}{\sqrt{\binom{n}{j}}}.
    \label{eq:conditions_state}
\end{eqnarray}

Those conditions imply that the vectors can be written as
\begin{eqnarray}
    \ket{v_0} &=& \sum_{j = 0}^{n} K \frac{p^*_j}{\sqrt{\binom{n}{j}}} \ket{s_j^n} + \ket{v_0^\perp},\\
    \ket{v_1} &=& \sum_{j = 0}^{n} K \frac{q^*_j}{\sqrt{\binom{n}{j}}} \ket{s_j^n} + \ket{v_1^\perp},
\end{eqnarray}
where $\ket{v_0^\perp}$ and $\ket{v_1^\perp}$ are two vectors, not necessarily orthogonal, that describe the component of the respective vector in the subspace orthogonal to the one generated by the vectors $\ket{s_j^n}$.
Now we want to apply a procedure similar to the Gram–Schmidt decomposition to the vectors $\ket{v_0^\perp}$ and $\ket{v_1^\perp}$. We define
\begin{eqnarray}
    \ket{\theta_0} &=& \frac{\ket{v_0^\perp}}{\sqrt{\braket{v_0^\perp}{v_0^\perp}}}, \\ 
    Kx &=& \sqrt{\braket{v_0^\perp}{v_0^\perp}}, \\
    Ky &=& \braket{\theta_0}{v_1^\perp}, \\ 
    \ket{\theta_1} &=& \frac{\ket{v_1^\perp}-\ket{\theta_0}\braket{\theta_0}{v_1^\perp}}{\sqrt{\braket{v_1^\perp}{v_1^\perp}-\braket{v_1^\perp}{\theta_0}\braket{\theta_0}{v_1^\perp}}}, \\ 
    Kw &=& \braket{\theta_1}{v_1^\perp}.
\end{eqnarray}

With those definitions, the two vectors $\ket{v_0}$ and $\ket{v_1}$ can be rewritten as
\begin{eqnarray}
    \ket{v_0} &=& \sum_{j = 0}^{n} K \frac{p^*_j}{\sqrt{\binom{n}{j}}} \ket{s_j^n} + K x \ket{\theta_0}, \\ 
    \ket{v_1} &=& \sum_{j = 0}^{n} K \frac{q^*_j}{\sqrt{\binom{n}{j}}} \ket{s_j^n} + K y \ket{\theta_0} + K w \ket{\theta_1},
\end{eqnarray}
which is the general form presented in the main text.





\section{Efficient base completion algorithm}

As expressed in the main text, we aim at finding an efficient algorithm that, given two orthonormal vectors $\ket{v_0}$ and $\ket{v_1}$, returns a unitary matrix $U$ that has as its first two columns the two vectors. The idea for the algorithm came from the procedure for the QR decomposition with the Householder reflections \cite{Numerical_recipes_2007}.

We define: 
\begin{eqnarray}
    \ket{x_0} &=& \ket{v_0}, \\ 
    \alpha_0 &=& -\frac{\braket{e_0}{x_0}}{\abs{\braket{e_0}{x_0}}}, \\ 
    \ket{u_0} &=& \ket{x_0}-\alpha_0 \ket{e_0}.
\end{eqnarray}
With that, we define a first unitary matrix as
\begin{equation}
    Q_0 = \mathbb{I}-2\frac{\ketbra{u_0}{u_0}}{\braket{u_0}{u_0}},
\end{equation}
where $\ket{e_0} = \ket{0}^{\otimes n}$ is the first element of the standard orthonormal basis. We then consider
\begin{eqnarray}
    \ket{x_1} &=& Q_0\ket{v_1}, \\ 
    \alpha_1 &=& -\frac{\braket{e_1}{x_1}}{\abs{\braket{e_1}{x_1}}}, \\ 
    \ket{u_1} &=& \ket{x_1}-\alpha_1 \ket{e_1},
\end{eqnarray}
where $\ket{e_1} = \ket{1} \otimes \ket{0}^{\otimes (n-1)}$ is the second element of the standard orthonormal basis. Then we define a second unitary matrix as
\begin{equation}
    Q_1 = \mathbb{I}-2\frac{\ketbra{u_1}{u_1}}{\braket{u_1}{u_1}}.
\end{equation}
Finally, we define a matrix as
\begin{equation}
    U' = Q_0 Q_1 \big(\mathbb{I}+(\alpha_0-1)\ketbra{e_0}{e_0}+(\alpha_1-1)\ketbra{e_1}{e_1}\big),
\end{equation}
which has the first two columns that coincide with the vectors $\ket{v_0}$ and $\ket{v_1}$.

As a note in the main text, we want the unitary matrix with the first two rows as the covectors $\bra{v_0}$ and $\bra{v_1}$, so we need the conjugate of that matrix $U = U'$.


\section{Trace operation in Quantum-to-quantum Bernoully factories}

\begin{figure}[t]
        \centering
        \[\Qcircuit @C=1em @R=1em {
        &&& \lstick{\ket{z}} & \qw & \multigate{9}{U} & \qw & \qw &\rstick{\ket{\psi_O}} \qw\\
        &&& \lstick{\ket{z}} & \qw & \ghost{U} & \qw & \meter & \rstick{q_2 = 0} \cw\\
        &&& \cdots& & \nghost{U} & & \cdots \\
        &&& \lstick{\ket{z}} & \qw & \ghost{U} & \qw & \meter & \rstick{q_n = 0} \cw\\
        &&& \lstick{\ket{0_c}} & \qw & \ghost{U} & \qw & \meter & \rstick{q_{n+1} = 0} \cw\\
        &&& \cdots& & \nghost{U} & & \cdots \\
        &&& \lstick{\ket{0_c}} & \qw & \ghost{U} & \qw & \meter & \rstick{q_{n+m} = 0} \cw\\
        &&& \lstick{\ket{z/0_c}} & \qw & \ghost{U} & \qw &  \ground \qw\\
        &&& \cdots& & \nghost{U} & & \cdots \\
        &&& \lstick{\ket{z/0_c}} & \qw & \ghost{U} & \qw &  \ground \qw\\
        {\inputgroupv{1}{4}{1em}{3.2 em}{n}} \\
        {\inputgroupv{5}{7}{1em}{2.1 em}{m}} \\
        {\inputgroupv{8}{10}{1em}{2.1 em}{r}} \\
        }\]
        
        \caption{\textbf{Quantum Bernoulli factory general circuit with traced qubits.} The considered circuit is the one presented in Fig.~1 of the main text with the addition of $r$ qubits, that can be in the coin state or in the computational state $\ket{0_c}$, which at the end of the protocol are traced.}
        \label{fig: BF_circuit_trace}
\end{figure}
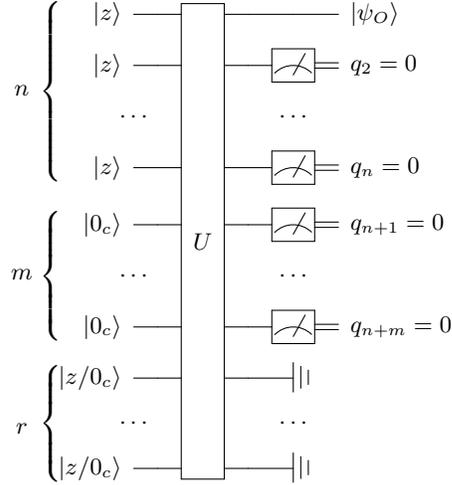

Now we address the task mentioned in the demonstration in the main text, namely that the trace operation is not necessary for the implementation of a Quantum-to-Quantum Bernoulli factory. In particular, we show that for fixed $m$ and $n$, the use of additional traced states does not increase the success probability.
If we consider the trace as a measurement where the outcome is discarded, we can see that the general circuit is still the one presented in Fig.~\ref{fig: BF_circuit_trace}. It can also be shown that all results for the set of simulatable functions and the number of resources required remain valid.

To show that the success probability is not optimal we consider the state $\ket{\psi_T}$ after the unitary evolution. For the Schmidt decomposition, we can write the state as
\begin{equation}
    \ket{\psi_T} = \sum_j \alpha_j \ket{\psi_j} \otimes \ket{\phi_j},
\end{equation}
where states $\ket{\psi_j}$ are the states of the $n+m$ qubits and $\ket{\phi_j}$ are states of the system that will be traced.
We now apply the measurement to the first system, obtaining the state
\begin{equation}
    \ket{\psi_M} = \frac{1}{\sqrt{\sum_k \abs{\alpha_k}^2 \abs{\beta_k}^2}}\sum_j \alpha_j \beta_j \ket{g_j(z)} \otimes \ket{\phi_j}.
\end{equation}
If a couple of indices $j$ and $j'$ where $\beta_j \neq 0$, $\beta_{j'} \neq 0$ and $\ket{g_j(z)} \neq \ket{g_{j'}(z)}$, then, after the trace operation, the resulting state becomes mixed. Since this possibility is excluded from the definition of the Quantum-to-Quantum Bernoulli factory, the only possibility is that all the terms with $\beta_j \neq 0$ are associated with the same state $\ket{g(z)}$.
Then the success probability of the protocol is
\begin{equation}
    \mathrm{Pr} = \sum_j \abs{\alpha_j}^2 \abs{\beta_j}^2.
\end{equation}
Each $\abs{\beta_j}^2$ is the probability of the projection of $\ket{\psi_j}$ over $\ket{g(z)}$, which corresponds to the success probability of a Quantum-to-Quantum Bernoulli factory with $n$ qubits, $m$ ancillary bits and without traced states. As previously shown, those probabilities are bounded above by the probability $\mathrm{Pr}_n$ of the optimal Bernoulli factory, and thus
\begin{equation}
    \mathrm{Pr} = \sum_j \abs{\alpha_j}^2 \abs{\beta_j}^2 \leq  \sum_j\abs{\alpha_j}^2 \mathrm{Pr}_n = \mathrm{Pr}_n.
\end{equation}
This inequality shows that adding the trace operation does not introduce any advantage in the implementation of a quantum-to-quantum Bernoulli factory.

\section{Multivariate Quantum-to-Quantum Bernoulli factory problem demonstration.}
In this Section, we present the demonstration for the Multivariate version of the problem.
As presented in the main text the general circuit is the one presented in Fig.~3 of the main text.
The input state of the circuit can be written as
\begin{equation}
\begin{split}
    \ket{\psi_I} &= \ket{0_c}^{\otimes m} \otimes \biggl(\frac{z_k\ket{0_c}+\ket{1_c}}{\sqrt{1+\abs{z}^2}}\biggr)^{\otimes n_k} \otimes \dots \otimes\biggl(\frac{z_1\ket{0_c}+\ket{1_c}}{\sqrt{1+\abs{z}^2}}\biggr)^{\otimes n_1}  \\ 
    &= \biggl(\frac{1}{1+\abs{z_1}^2}\biggr)^{\frac{n_1}{2}} \dots \biggl(\frac{1}{1+\abs{z_k}^2}\biggr)^{\frac{n_k}{2}}\sum_{j_1=0 \dots j_k=0}^{n_1 \dots n_k} z_1^{j_1} \dots z_k^{j_k} \sqrt{\binom{n_1}{j_1} \dots \binom{n_k}{j_k}} \ket{0_c}^{\otimes m} \otimes  \ket{s_{j_k}^{n_k}} \otimes \dots \otimes \ket{s_{j_1}^{n_1}}.
\end{split}
\end{equation}
As for the single parameter case, we note that the coefficients that compose the state are all the multivariate monomials with degree at most $n_j$ for each variable. This implies that $\text{Rat}(z_1, \dots, z_k)$ is a superset of the simulable functions and that if exist a circuit implementing a function $s(z_1,\dots, z_k)$ then the required number of qubits for each variable is at least $n_j \geq \deg_j(s(z_1, \dots, z_k))$.

For a function $s(z_1, \dots, z_k) = \frac{P(z_1, \dots, z_k)}{Q(z_1, \dots, z_k)}$ we define
\begin{eqnarray}
    P(z_1, \dots, z_k) &=& \sum_{j_1=0 \dots j_k=0}^{n_1 \dots n_k} p_{j_1 \dots j_k} z_1^{j_1} \dots z_k^{j_k}, \\ 
    Q(z_1, \dots, z_k) &=& \sum_{j_1=0 \dots j_k=0}^{n_1 \dots n_k} q_{j_1 \dots j_k} z_1^{j_1} \dots z_k^{j_k}.
\end{eqnarray}
Following a similar approach to the one of the standard version of the problem, we define the vectors $\ket{v_0}$ and $\ket{v_1}$ associated with the first two columns of the unitary transformation as
\begin{eqnarray}
    \ket{v_0} &=& \sum_{j_1 = 0 \dots j_k = 0}^{n_1 \dots n_k} K \frac{p^*_{j_1 \dots j_k}}{\sqrt{\binom{n_1}{j_1} \dots \binom{n_k}{j_k}}} \ket{0_c}^{\otimes m} \otimes\ket{s_{j_k}^{n_k}} \otimes \dots \otimes \ket{s_{j_1}^{n_1}} + K x \ket{\theta_0}, \\ 
    \ket{v_1} &=& \sum_{j_1 = 0 \dots j_k = 0}^{n_1 \dots n_k} K \frac{q^*_{j_1 \dots j_k}}{\sqrt{\binom{n_1}{j_1} \dots \binom{n_k}{j_k}}} \ket{0_c}^{\otimes m} \otimes\ket{s_{j_k}^{n_k}} \otimes \dots \otimes \ket{s_{j_1}^{n_1}} + K y \ket{\theta_0} + K w \ket{\theta_1}.
\end{eqnarray}
If we then define the quantities
\begin{eqnarray}
    a &=& \sum_{j_1 = 0 \dots j_k = 0}^{n_1 \dots n_k} \frac{\abs{q_{j_1 \dots j_k}}^2}{\binom{n_1}{j_1} \dots \binom{n_k}{j_k}}, \\
    b &=& \sum_{j_1 = 0 \dots j_k = 0}^{n_1 \dots n_k} \frac{\abs{p_{j_1 \dots j_k}}^2}{\binom{n_1}{j_1} \dots \binom{n_k}{j_k}}, \\ 
    c &=& \sum_{j_1 = 0 \dots j_k = 0}^{n_1 \dots n_k} \frac{p_{j_1 \dots j_k}q_{j_1 \dots j_k}}{\binom{n_1}{j_1} \dots \binom{n_k}{j_k}},
\end{eqnarray}
then the solutions for the variables $x$ and $y$ of the orthonormality equations are the same as the ones found for the single-parameter case.
The existence of the vector $\ket{\theta_0}$ orthogonal to all the vectors $\ket{0_c}^{\otimes m} \otimes\ket{s_{j_k}^{n_k}} \otimes \dots \otimes \ket{s_{j_1}^{n_1}}$ requires that $2^{n_1+\dots+n_k} > (n_1+1)\dots (n_k+1)$ , a condition that is always satisfied except for the case $\forall j \; n_j = 1$ where an ancillary qubit is required.
This implies that each of the multivariate rational functions $s(z_1\dots z_k)$ is simulable and that the required number of qubits corresponds to the bound $n_j = \deg_j(s(z_1, \dots, z_k))$ without needing ancillary qubits except in the case with $n_j = 1$.
Similarly to the single parameter case, if $a = b$ and $c=0$ no ancillary qubits are required.

For completeness, the optimal success probability for this scenario is found to be:
\begin{equation}
    Pr_n = \frac{2\biggl(\abs{\sum_{j_1 = 0 \dots j_k = 0}^{n_1 \dots n_k} p_{j_1 \dots j_k} z_1^{j_1} \dots z_k^{j_k}}^2+\abs{\sum_{j_1 = 0 \dots j_k = 0}^{n_1 \dots n_k} q_{j_1 \dots j_k} z_1^{j_1} \dots z_k^{j_k}}^2\biggr)}{\biggl(1+\abs{z_1}^2\biggr)^{n_1} \dots \biggl(1+\abs{z_k}^2\biggr)^{n_k} \biggl[\sqrt{\biggl(a-b \biggr)^2+4\abs{c}^2}+a+b\biggr]}.
    \label{eq:succ_prob_gen}
\end{equation}


\section{Multifunctional Quantum-to-quantum Bernoulli factory}
In this Section, we show that given two simulable functions $g_0(z) = P(z)/Q(z)$ and $g_1(z) = R(z)/S(z)$, there always exists a unitary matrix that implements the corresponding Multifunctional Quantum-to-quantum Bernoulli factory. Before proceeding with the demonstration, we first define and prove an extension of the Unitary dilation theorem \cite{Levy2014}.

\textbf{Unitary dilation theorem:} For any contraction matrix $A \in \mathbb{C}^{N \times M}$ (i.e. $\norm{A}_2\leq1$) there exists an unitary matrix $U\in \text{U}(N+M)$ that contains $A$ as a submatrix. That matrix is called a dilation of $A$.
Moreover all the dilations $U$ are equivalent \footnote{In this context, the equivalence relation is defined as: $U\sim V$ if there exist two unitary matrices $U_1, U_2 \in U(N)$ such that $V = (\mathds{I}\oplus U_1)U(\mathds{I}\oplus U_2)$} to the matrix
\begin{equation}
    U = \begin{pmatrix}
        A & \sqrt{\mathds{I}_N - AA^\dagger}\\
        \sqrt{\mathds{I}_M - A^\dagger A} & -A^\dagger
    \end{pmatrix}.
    \label{eq:dilation}
\end{equation}

\textit{Proof.} We now prove the Unitary dilation theorem for rectangular matrices, where in particular we need to prove that the matrix $U$ defined in Eq.~\eqref{eq:dilation} is unitary. First, we use the singular value decomposition and we rewrite the contraction as $A = U\Sigma V^\dagger$ where $U$ and $V$ are respectivly an $N\times N$ and an $M\times M$ unitary matrix, and $\Sigma$ is an $N \times M$ diagonal matrix with positive entries all less than one.
\begin{equation}
    UU^\dagger = \begin{pmatrix}
        AA^\dagger+\sqrt{\mathds{I}_N-AA^\dagger}\sqrt{\mathds{I}_N-AA^\dagger}^\dagger & A\sqrt{\mathds{I}_M-A^\dagger A}^\dagger-\sqrt{\mathds{I}_N-AA^\dagger }A\\
        \sqrt{\mathds{I}_M-A^\dagger A}A^\dagger-A^\dagger\sqrt{\mathds{I}_N-AA^\dagger}^\dagger & A^\dagger A + \sqrt{\mathds{I}_M-A^\dagger A}\sqrt{\mathds{I}_M-A^\dagger A}^\dagger
    \end{pmatrix}.
\end{equation}
For the diagonal component we have 
\begin{equation}
\begin{aligned}
    AA^\dagger+\sqrt{\mathds{I}_N-AA^\dagger}\sqrt{\mathds{I}_N-AA^\dagger}^\dagger &= U\Sigma \Sigma ^\dagger U^\dagger + \sqrt{\mathds{I}_M-U\Sigma \Sigma ^\dagger U^\dagger} \sqrt{\mathds{I}_M - U\Sigma \Sigma ^\dagger U^\dagger}^\dagger \\
    &= U\Sigma \Sigma ^\dagger U^\dagger + U\sqrt{\mathds{I}_M-\Sigma \Sigma ^\dagger }\sqrt{\mathds{I}_M - \Sigma \Sigma ^\dagger}^\dagger U^\dagger\\
    &= U\Sigma \Sigma ^\dagger U^\dagger + U\sqrt{\mathds{I}_M-\Sigma \Sigma ^\dagger }\sqrt{\mathds{I}_M - \Sigma \Sigma ^\dagger} U^\dagger\\ 
    &= U\left (\Sigma \Sigma ^\dagger + \mathds{I}_M-\Sigma \Sigma ^\dagger  \right)U^\dagger\\
    &= \mathds{I}_M,
\end{aligned}
\end{equation}
and similarly
\begin{equation}
    A^\dagger A+ \sqrt{\mathds{I}_M-A^\dagger A}\sqrt{\mathds{I}_M-A^\dagger A}^\dagger = \mathds{I}_N.
\end{equation}
Instead, for the off-diagonal component, we have 
\begin{equation}
\begin{aligned}
    A\sqrt{\mathds{I}_M-A^\dagger A}^\dagger-\sqrt{\mathds{I}_N-AA^\dagger }A &= U\Sigma V^\dagger V\sqrt{\mathds{I}_M-\Sigma^\dagger \Sigma}^\dagger V^\dagger-U \sqrt{\mathds{I}_N-\Sigma \Sigma^\dagger} U^\dagger U \Sigma V^\dagger \\
    &= U\left(\Sigma  \sqrt{\mathds{I}_M-\Sigma^\dagger \Sigma} -\sqrt{\mathds{I}_N-\Sigma \Sigma^\dagger} \Sigma\right)V^\dagger.
\end{aligned}
\end{equation}
Here we can consider two cases, If $N\geq M$ then the matrix $\Sigma$ is in the form
\begin{equation}
    \Sigma = \begin{pmatrix}
        D\\0
    \end{pmatrix},
\end{equation}
where $D$ is the $M\times M$ diagonal component of the singular values and $0$ is the $N-M \times M$ component with all the entry equal to $0$. Hence, we find that: 
\begin{equation}
\begin{aligned}
    \Sigma  \sqrt{\mathds{I}_M-\Sigma^\dagger \Sigma} -\sqrt{\mathds{I}_N-\Sigma \Sigma^\dagger} \Sigma &= \begin{pmatrix} D\\0 \end{pmatrix}\sqrt{\begin{pmatrix}\mathds{I}_M\end{pmatrix}-\begin{pmatrix} D^* & 0\end{pmatrix}\begin{pmatrix}D\\0\end{pmatrix}}-\sqrt{\begin{pmatrix}\mathds{I}_M & 0 \\ 0 & \mathds{I}_{N-M} \end{pmatrix}-\begin{pmatrix}D\\0\end{pmatrix}\begin{pmatrix}D^*&0\end{pmatrix}}\begin{pmatrix}D\\0
    \end{pmatrix}\\
    &= \begin{pmatrix}D\\0\end{pmatrix} \begin{pmatrix}\sqrt{\mathds{I}_M-D^*D}\end{pmatrix}- \begin{pmatrix} \sqrt{\mathds{I}_M-D^*D} &0 \\ 0 & \mathds{I}_{N-M}\end{pmatrix}\begin{pmatrix}D\\0\end{pmatrix}\\
    &= \begin{pmatrix}D\sqrt{\mathds{I}_M-D^*D}-\sqrt{\mathds{I}_M-D^*D}D \\ 0\end{pmatrix}\\
    & = \begin{pmatrix} 0 \\ 0\end{pmatrix}
\end{aligned}.
\end{equation}
Conversely, if $N\geq M$ then $\Sigma$ is in the form
\begin{equation}
    \Sigma = \begin{pmatrix}
        D & 0
    \end{pmatrix},
\end{equation}
and the demonstration proceeds analogously to the other case.

An analogous demonstration can then be performed for the other off-diagonal element. Thus, we obtain the final result
\begin{equation}
    UU^\dagger = \begin{pmatrix}
    \mathds{I}_M & 0\\
    0 & \mathds{I}_N
\end{pmatrix},
\end{equation}
which means that $U$ is unitary as we want to demonstrate. $\square$

We can now proceed with showing that given two simulable functions, there always exists a unitary matrix that implements the corresponding Multifunctional Quantum-to-quantum Bernoulli Factory. First, we define the following four vectors
\begin{eqnarray}
    \ket{p} &=& \sum_{j = 0}^{n} K \frac{p^*_j}{\sqrt{\binom{n}{j}}} \ket{s_j^n}, \\
    \ket{q} &=& \sum_{j = 0}^{n} K \frac{q^*_j}{\sqrt{\binom{n}{j}}} \ket{s_j^n}, \\
    \ket{r} &=& \sum_{j = 0}^{n} H \frac{r^*_j}{\sqrt{\binom{n}{j}}} \ket{s_j^n}, \\ 
    \ket{s} &=& \sum_{j = 0}^{n} H \frac{s^*_j}{\sqrt{\binom{n}{j}}} \ket{s_j^n},
\end{eqnarray}
and the rectangular matrix
\begin{equation}
    M = \ketbra{00}{p}+\ketbra{10}{q}+\ketbra{10}{r}+\ketbra{11}{s}.
\end{equation}
With those definitions, the matrix $M$ should be a submatrix of the unitary matrix that implements the Multifunctional Quantum-to-quantum Bernoulli factory up to a scalar factor.
If we now define
\begin{equation}
    A = \frac{M}{r\norm{M}_2},
\end{equation}
where $\norm{M}_2$ is the Spectral norm and $r\geq1$ is a scalar factor, such that $A$ is a proper contraction, then we can construct the unitary matrix implementing the Multifunctional Quantum-to-quantum Bernoulli factory associated with the functions $g_0(z)$ and $g_1(z)$ using the dilation theorem.


%